\UseRawInputEncoding
\pdfoutput=1

\documentclass[12pt]{article}

\usepackage{amsfonts,eurosym,geometry,ulem,graphicx,caption,color,setspace,sectsty,footmisc,caption,natbib,pdflscape,hyperref}
\usepackage[utf8]{inputenc}
\usepackage{amssymb,amsmath,amsthm} 
\usepackage{mathrsfs}
\usepackage{mathtools}
\usepackage{bm}
\usepackage{bbm}
\usepackage{nicefrac}
\usepackage{array}
\usepackage{tabularx}
\usepackage{url}
\usepackage{multirow}
\usepackage{makecell}
\usepackage{accents}
\usepackage{doi}
\usepackage{float}
\usepackage{hhline}

\usepackage{natbib}

\usepackage{changes}

\usepackage{graphicx} 
\graphicspath{{Figures/}} 
\usepackage{caption}
\usepackage{subcaption}
\usepackage{adjustbox}

\usepackage{enumitem} 

\usepackage{varioref} 
\usepackage[toc,page]{appendix}

\newcommand{\mb}{\bm{m}}
\newcommand{\cb}{\bm{c}}
\newcommand{\kb}{\bm{k}}
\newcommand{\zb}{\bm{z}}
\newcommand{\Mb}{\bm{M}}

\newcommand{\ub}{\bm{u}}

\newcommand{\Ub}{\bm{U}}
\newcommand{\vb}{\bm{v}}
\newcommand{\xb}{\bm{x}}
\newcommand{\gb}{\bm{g}}

\newcommand{\Ex}{\textsf{E}}
\newcommand{\eq}{\,=\,}
\newcommand{\var}{\textsf{var}}
\newcommand{\cov}{\textsf{cov}}
\newcommand{\etab}{\bm{\eta}}

\newcommand{\gammab}{\bm{\gamma}}
\newcommand{\bphi}{\bm{\phi}}
\newcommand{\hbphi}{\hat{\bm{\phi}}}
\newcommand{\heta}{\hat{\eta}}
\newcommand{\ff}{\textsf{f}}
\newcommand{\dd}{\textsf{d}}
\newcommand{\FF}{\textsf{F}}
\newcommand{\PP}{\textsf{P}}

\newcommand{\CC}{\textsf{C}}
\newcommand{\inn}{\,\in\,}
\newcommand{\ijn}{1 \leq i < j \leq n}
\newcommand{\lleq}{\,\leq\,}
\newcommand{\eps}{\varepsilon}
\newcommand{\nn}{\nonumber}
\DeclareMathOperator*{\ssup}{\textsf{sup}}

\DeclareMathOperator*{\mmax}{\textsf{max}}
\DeclareMathOperator*{\argmmax}{\textsf{arg\,max}}
\newtheorem{remark}{Remark}

\newcommand{\R}{\mathbb{R}}

\newcommand{\e}{\varepsilon}

\newcommand\norm[1]{\lVert#1\rVert}

\newtagform{scroman}[]()

\newcommand*\samethanks[1][\value{footnote}]{\footnotemark[#1]}

\DeclareMathOperator*{\argmin}{arg\,min}

\DeclarePairedDelimiter\floor{\lfloor}{\rfloor}
\newtheoremstyle{break}
  {\topsep}{\topsep}%
  {\itshape}{}%
  {\bfseries}{}%
  {\newline}{}%

\theoremstyle{definition} 

\theoremstyle{plain} 

\theoremstyle{break}
\newtheorem{assumption}{Assumption}
\newtheorem{proposition}{Proposition}
\newtheorem{lemma}{Lemma}

\theoremstyle{remark} 

\newcolumntype{L}[1]{>{\raggedright\let\newline\\arraybackslash\hspace{0pt}}m{#1}}
\newcolumntype{C}[1]{>{\centering\let\newline\\arraybackslash\hspace{0pt}}m{#1}}
\newcolumntype{R}[1]{>{\raggedleft\let\newline\\arraybackslash\hspace{0pt}}m{#1}}

\begin{document}

\begin{titlepage}
\title{Consistent Estimation of Multiple Breakpoints in Dependence Measures}
\author{Marvin Borsch\thanks{Institute of Econometrics and Statistics, University of Cologne. Financial support by Deutsche Forschungsgemeinschaft (DFG grant ‘Strukturbr\"uche und Zeitvariation in hochdimensionalen Abh\"angigkeitsstrukturen’) is gratefully acknowledged.} \and Alexander Mayer\samethanks \and Dominik Wied\samethanks}
\date{\today}
\maketitle
\begin{abstract}
\noindent This paper proposes different methods to consistently detect multiple breaks in copula-based dependence measures, mainly focusing on Spearman's $\rho$. The leading model is a factor copula model due to its usefulness for analyzing data in high dimensions. Starting with the classical binary segmentation, also the more recent wild binary segmentation (WBS) and a procedure based on an information criterion are considered. For all procedures, consistency of the estimators for the location of the breakpoints as well as the number of breaks is proved. Monte Carlo simulations indicate that WBS performs best in many, but not in all, situations. A real data application on recent Euro Stoxx 50 data reveals the usefulness of the procedures.\\ 
\vspace{0in}

\bigskip
\end{abstract}
\setcounter{page}{0}
\thispagestyle{empty}
\end{titlepage}
\pagebreak \newpage

\doublespacing

\section{Introduction}

For asset allocation in financial markets and risk management purposes, dependence measures are of great interest. For example, it is necessary to estimate the variance-covariance matrix of asset returns to construct mean-variance efficient portfolios as introduced by \cite{markowitz1978portfolio}. In recent years, also copula-based/non-linear dependencies \citet[Chapter~5]{nelsen2007introduction} became popular, e.g. using a conditional multivariate version of Spearman's $\rho$ as in \cite{schmid2007multivariate} or \cite{penzer2012measuring}.
Therefore, estimators of these measures are crucial. However, in financial time series, breaks in dependence measures can occur. These happen quite frequently in times of crisis, e.g. during the financial crisis in 2008 or more recently in the beginning of 2020 when the Corona pandemic began. This phenomenon is generally known as the diversification meltdown. Hence, consistent estimation of the location of those break points as well as their number need to be considered in order to correctly estimate the dependencies between assets (in between the breaks).

In this paper, we propose methods for dating multiple breaks in copula-based dependence measures. The methods are broadly applicable, but our leading example is the factor copula framework as introduced in \cite{oh2017modeling}. This is especially useful in high-dimensional applications due to the sparse amount of parameters in the factor structure. Abstaining from a multivariate Gaussian distribution assumption, it can capture tail risk as well as the leverage effect. Breakpoint detection in a factor copula framework has already discussed been discussed by \cite{manner2019testing}. We extend their so-called `moment-based' test  to the detection of multiple breakpoints in copula-based dependence measures for filtered data. In doing so, we use recent results by \cite{nasri2022change} to derive consistency of our procedure for the location as well as the number of breaks from primitive conditions. 

The literature for structural changes in dependence measures is wide. \cite{wied2012testing}, for instance, test for breaks in the correlation using an extended functional delta method, while \cite{aue2009break} test for changes in the covariance matrix. Nonlinear dependence measures like Spearman's \(\rho\) are analyzed with respect to break points in \cite{gombay1999change}, \cite{wied2014fluctuation}, \cite{kojadinovic2016testing}, \cite{manner2019testing}, or \cite{stark2020testing}, among others.  Similar to \cite{wied2014fluctuation}, we use a CUSUM-type statistic but allow for multiple copula-based dependence measures including, for example, Spearman's \(\rho\), Gini's \(\gamma\), or Spearman's footrule. However, we follow suggestions by \cite{bucher2014detecting} for calculating ranks sequentially to improve power. Moreover, our test statistic is based on filtered return data instead of observable time series; an idea also recently employed by \cite{barassi2020change} to analyze changes in the conditional correlation.

In order to detect multiple breaks, we use a binary segmentation (BS) algorithm that goes back to \cite{vostrikova1981detecting} and which since then has been adopted frequently in the literature; see, e.g. \cite{bai1997estimating}. In principle, we follow the algorithm used in \cite{galeano2010shifts} or in \cite{galeano2014multiple}. More specifically, the maximum of a CUSUM-statistic within the subintervals is compared to some critical value. Here, and similar to \citet[Section 4.3]{bai1998estimating}, we derive the asymptotic distribution of the maximum statistic under the null hypothesis that there is no \textit{additional} break to take multiple testing into account. We also extend our results to the wild binary segmentation (WBS) algorithm of \cite{fryzlewicz2014wild}, which improves the detection of change points, especially for multiple breaks in close proximity.

The rest of this paper is structured as follows. Section \ref{sec:breaks and theory} contains the copula-based dependence measures, segmentation algorithms and the consistency results. In Section \ref{sec: Monte Carlo Simulation} Monte Carlo simulations are provided. Section \ref{sec: Empirical Application} gives an empirical application, while Section \ref{sec: Conclusion} concludes.

\section{Finding Multiple Breaks in Dependence Measures}\label{sec:breaks and theory}

\subsection{Modeling Breaks in Dependence Measures}
Suppose we observe a sample $\bm{Y}_1,\dots, \bm{Y}_T$ of length $T$ with a cross section $\bm{Y}_t \coloneqq (Y_{1,t},\dots,Y_{n,t})'$ of $n$ financial assets so that each $n \times 1$ vector $\bm{Y}_t$, $t \in \{1,\dots,T\}$, obeys a parametric location-scale specification
\begin{align} \label{eq: Data}
\bm{Y}_t = \bm{\mu}_t(\bm{\phi}^0) + \bm{\sigma}_t(\bm{\phi}^0) \bm{\eta}_t, 
\end{align}
where \(\etab_t \coloneqq (\eta_{1,t},\dots,\eta_{n,T})'\) are innovations with \(\Ex[\eta_{i,t}] = 0\) and \(\var[\eta_{i,t}] = 1\) for any \(i \in \{1,\dots,n\}\). Similar to \cite{chen2006estimation}, \cite{oh2013simulated}, and \cite{oh2017modeling}, we assume that the conditional mean and variance functions---respectively given by $\bm{\mu}_t(\bm{\phi}^0) \coloneqq (\mu_{1,t}(\bm{\phi}^0),\dots,\mu_{n,t}(\bm{\phi}^0))'$ and $\bm{\sigma}_t(\bm{\phi}^0) \coloneqq \textsf{diag}\{\sigma_{1,t}(\bm{\phi}^0),\dots,\sigma_{n,t}(\bm{\phi}^0)\}$---are (a) parametrically known up to a finite dimensional parameter $\bm{\phi}^0$ and (b)  measurable with respect to $\mathcal{F}_{t-1} \coloneqq \sigma(\{\bm{Y}_j,\, j < t\})$. The innovations \(\bm{\eta}_t\), on the other hand, are jointly independent of \(\mathcal{F}_{t-1}\). This setting allows for many AR-GARCH specifications commonly encountered in financial econometrics.

Assuming continuous margins \(\FF_{i,t}(x) \coloneqq \PP(\eta_{i,t} \leq x)\), \(x \in \R\), we can rephrase the conditional joint distribution as
\begin{equation}\label{eq:cop_decomp}
\bm{Y}_t \mid \mathcal{F}_{t-1} \sim \CC_t(\FF_{1,t}(\eta_{1,t}),\dots,\FF_{n,t}(\eta_{n,t})),
\end{equation}
where the copula \(\CC_{t}(\ub) = \PP(\Ub_{t} \leq \ub),\) \(\ub \in [0,1]^n\), \(\Ub_{t} \coloneqq (U_{1,t},\dots,U_{n,t})'\) with \(U_{i,t} \coloneqq \FF_{i,t}(\eta_{i,t})\), captures uniquely the dependence among the \(n\) variates in \(\etab_t\); see \cite{patton2006modelling}. In what follows, we aim at finding change points in certain dependence measures that can be solely expressed in terms of the copula. As argued, for example, in \cite{bucher2014detecting} or \cite{kojadinovic2016testing}, classical nonparametric tests based on sequential empirical processes have little power against alternatives that leave the margins unchanged and only involve a change in the copula. 

More specifically, for a finite collection of disjoint sets  \(\{\mathcal{G}_1,\dots,\mathcal{G}_G\}\) partitioning the cross-sectional index set  \(\{1,\dots,n\}\) and a collection of suitable bivariate functions \(h_l: [0,1]^2 \rightarrow \mathbb{R}\), \(l \in \{1,\dots,H\}\), we focus on averaged rank-based dependence measures of the type 
\begin{equation}\label{average-m}
m_{t,g,l} \coloneqq \frac{1}{\displaystyle{|\mathcal{G}_g| \choose 2}}\mathop{\sum}\limits_{\substack{\ijn\\i,j \inn \mathcal{G}_g}}m_{i,j,t,l},\;m_{i,j,t,l} \coloneqq \Ex[h_l(U_{i,t},U_{j,t})],\,g \in \{1,\dots,G\}. 
\end{equation}
We aim at locating and estimating change points of the \(p \times 1\) vector 
\[
\mb_t \coloneqq (\mb_{t,1}',\dots,\mb_{t,G}')',\;\mb_{t,g} \coloneqq (m_{t,g,1},\dots,m_{t,g,H})',\;\;p \coloneqq GH.
\]
To illustrate, Spearman's \(\rho\), Spearman's footrule (\(\varrho\)), and Gini's \(\gamma\) fit into this setup with \(h_\rho(u,v) = 12uv-3\), \(h_\varrho(u,v) = 1-3|u-v|\), and \(h_\gamma(u,v) = 2(|u+v-1|-|u-v|)\) for \((u,v) \in [0,1]^2\), respectively; see, e.g., \cite{nelsen2007introduction}. The group structure is often used in empirical studies, where some degree of homogeneity within each group can be expected so that \(m_{i,j,t,l} = m_{g,t,l}\) for all \((i,j) \in \mathcal{G}_g\). Group structures are commonly used in empirical work; see, e.g. \cite{oh2017modeling}, \cite{opschoor2021closed}, or \cite{oh2021dynamic}.
Averaged dependence measures are used frequently and have been found to perform well; see, e.g.  \cite{schmid2007multivariate}, \cite{quessy2009theoretical}, \cite{kojadinovic2016testing}, or \cite{manner2019testing}.

Importantly, the above-mentioned dependence measures do not depend on the univariate margins, i.e.
\begin{equation}\label{eq: dmeasure_true}
m_{i,j,t,l} = \int_{[0,1]^2}h_l(u,v)\,\textsf{d}\CC_{i,j,t}(u,v),
\end{equation}
where \(\CC_{i,j,t}(u_i,u_j) \coloneqq \CC_t(\ub^{(i,j}))\), with \(\ub^{(i,j)} = (1,\dots,1,u_i,1,\dots,1,u_j,1\dots,1)'\) for some \(n \times 1\) vector \(\ub \in [0,1]^n\), denotes the bivariate marginal copula of \(\CC_{t}\). A change in \(\mb_{t}\) over time must thus be due to a change in \(\CC_{t}\). We therefore consider the following break-point scenario as summarized by Assumption \ref{A-break}.

\renewcommand{\theassumption}{A}
\begin{assumption}\label{A-break} The following holds true:
\begin{enumerate}[label= \textnormal{(A\arabic*)}]
\item\label{A-break1} Consider a finite partition \(0 \eqqcolon z_{0}^0 < z_{1}^0 < \dots < z_{\ell^0}^0 < z_{\ell^0 +1}^0 \coloneqq 1\). If \(\floor{z_{i-1}^0T} \leq \floor{t/T} \leq \floor{z_{i}^0T}\), \(i \in \{1,\dots,\ell^0+1\}\),
 then
\(\normalfont\CC_{t}(\ub)= \CC_{i}(\ub)\), where  \(\normalfont\CC_{1},\dots, \CC_{\ell^0+1}\) denote copulae, which are pairwise different in at least one point \(\ub \in [0, 1]^n\) such that \(\mb_{t} = \gb(t/T)\), where \(\gb: [0,1] \rightarrow \mathbb{R}^p\) is a vector-valued step function defined as
\[
\gb(z) = \sum_{i \eq 0}^{\ell^0} \gammab_{i}^0 1\{z \inn [z_{i}^0,z_{i+1}^0)\},\,\, \gb(1) \coloneqq \gammab_{\ell^0}^0.
\]
The \(p \times 1\) level vectors \(\gammab_{0}^0,\dots,\gammab_{\ell^0}^0\), as well as \(z_{1}^0,\dots,z_{\ell^0}^0\) and \(\ell^0\) are finite constants independent of \(T\).
\item\label{A-break2} The partial derivatives \(\normalfont\dot{\CC}_{i,k+1} \coloneqq \partial\CC_{k+1}/\partial{u_i}\) exist and are continuous on \(\{\ub \in [0,1]^n: 0 < u_i < 1\}\),  for any \(k \in \{0,1,\dots,\ell^0\}\) and \(i \in \{1,\dots,n\}\).
\item\label{A-break3} \(\etab_1,\dots, \etab_T\) is a sample from a process \(\{\etab_t\}_{t \inn \mathbb{Z}}\) of independently distributed \(n \times 1\) random vectors \(\etab_t = (\eta_{1,t},\dots,\eta_{n,t})'\) with strictly stationary and continuous univariate margins \(\normalfont\eta_{i,t} \sim \FF_i\), \(i \in \{1,\dots,n\}\). 
\end{enumerate}
\end{assumption}

The function $\gb(\cdot)$ defined in part \ref{A-break1} of Assumption \ref{A-break} specifies the timing and the size of the changes in the dependence measures. This specification is tailored towards abrupt changes and allows for a change in only a subset of $\mb_t$; see \cite{galeano2014multiple, galeano2017dating} for a detailed discussion. Part \ref{A-break2} is due to \cite{segers2012smooth} and imposes a smoothness condition on the partial derivatives of the copula. Similar to \cite{bucher2014detecting} or \cite{kojadinovic2016testing}, by imposing part \ref{A-break3} of Assumption \ref{A-break}, we maintain stationary marginal distributions, i.e. the joint distribution is only affected by a break in the copula. The following remark illustrates the setting within the context of factor copulas. 

\begin{remark}\label{remark:factor} 
Recently, factor copula models gained some popularity as a parsimonious yet flexible way to model the cross-sectional dependence among a possibly large number of financial assets; see, e.g. \textnormal{\cite{oh2013simulated}}, \textnormal{\cite{krupskii2013factor}}, \textnormal{\cite{creal2015high}}, or \textnormal{\cite{opschoor2021closed}}. According to this literature, one may assume that the unknown copula \(\normalfont\CC_t\) governing Eq. \textnormal{\eqref{eq:cop_decomp}} can be generated from an auxiliary factor model of the form \(X_{i,t} = \beta_{i,t}Z_t + \eps_{i,t}\), where  \(Z_t\), \(\beta_{i,t}\), and \(\eps_{i,t}\) denote factor, factor loading, and idiosyncratic component, respectively. For example, the group structure mentioned above arises naturally within the `block-equidependence' model from \textnormal{\cite{oh2017modeling}} where assets are driven by some latent market factor with different intensity in each group/industry resulting in equal dependence between the assets in each group but different dependence across groups. Breaks in dependence measures could result from structural changes in loadings, i.e. if \(\beta_{i,t} \neq \beta_{i,t+1}\) for some \(\normalfont t \in \{1,\dots,T\}\); an issue investigated below in more detail.
\end{remark}

\subsection{Segmentation Algorithms}
Suppose \(\hbphi\) is a  \(\sqrt{T}\)-consistent estimator of \(\bm{\phi}^0\) so that we can define the residuals \(\heta_{i,t} \coloneqq \sigma_{i,t}^{-1}(\hbphi)(Y_{i,t}-\mu_{i,t}(\hbphi))\). Moreover, for any \(1 \leq k< m\leq T\), let \(\hat{U}_{i,t}^{k:m}\), \(i \in \{1,\dots,n\}\), \(t \in \{k,\dots,m\}\), denote the rank of \(\heta_{i,t}\) among \(\heta_{i,k},\dots,\heta_{i,m}\). We then define for any \([a,b] \subseteq [0,1]\), a sample analogue of \eqref{average-m} by   
\begin{equation}\label{s-average-m}
\hat{m}_{g,l}^{\floor{aT}+1:\floor{bT}}  \coloneqq \frac{1}{\displaystyle{|\mathcal{G}_g| \choose 2}}\mathop{\sum}\limits_{\substack{\ijn\\i,j \inn \mathcal{G}_g}}\hat{m}_{i,j,l}^{\floor{aT}+1:\floor{bT}},\, l \in \{1,\dots,H\},\, g \in \{1,\dots,G\},
\end{equation}
where 
\[
\hat{m}_{i,j,l}^{\floor{aT}+1:\floor{bT}} \coloneqq \frac{1}{\floor{bT}-\floor{aT}}\sum_{t \eq \floor{aT}+1}^{\floor{bT}}h_l(\hat{U}_{i,t}^{\floor{aT}+1:\floor{bT}},\hat{U}_{j,t}^{\floor{aT}+1:\floor{bT}}),
\]
so that we obtain the \(p \times 1\) vector \(\hat{\mb}^{\floor{aT}+1:\floor{bT}} \coloneqq (\hat{\mb}_{1}^{\floor{aT}+1:\floor{bT}}{}',\dots,\hat{\mb}_{G}^{\floor{aT}+1:\floor{bT}}{}')'\), where \(\hat{\mb}_{g}^{\floor{aT}+1:\floor{bT}} \coloneqq (\hat{m}_{g,1}^{\floor{aT}+1:\floor{bT}},\dots,\hat{m}_{g,H}^{\floor{aT}+1:\floor{bT}})'\). Next, define for a given interval \([a,b] \subseteq [0,1]\) with \(0\leq \floor{aT} < \floor{bT} \leq T\), the detector
\[
\hat{M}(a,b;z) = \, \frac{\floor{zT}-\floor{aT}}{\sqrt{\floor{bT}-\floor{aT}}}\left\Vert\hat{\mb}^{\floor{aT}+1:\floor{zT}}-\hat{\mb}^{\floor{aT}+1:\floor{bT}}\right\Vert_2,
\]
and the fluctuation test statistic
\begin{equation}\label{eq:fluct_stat}
\hat{M}(a,b) \coloneqq \ssup_{z \inn [a,b]} \hat{M}(a,b;z) = \mmax_{z \inn \Pi_{T}(a,b)}\hat{M}(a,b;z),
\end{equation}
where \(\Pi_{T}(a,b)  \coloneqq \{\floor{aT}/T,(\floor{aT}+1)/T,\dots,(\floor{bT}-1)/T,\floor{bT}/T\}\), \(\Vert\cdot\Vert_2\) is the Euclidean norm, and the conventions \(\hat{M}(a,b;z) = 0\) if \(z = a\) and \(\Pi_{T} \coloneqq \Pi_{T}(0,1)\) are used.

Equation \eqref{eq:fluct_stat} can be related to the test statistic used in \cite{wied2014fluctuation} up to a scaling factor for $h = h_{\rho}$ and $H=1$. However, they calculate the ranks for Spearman's $\rho$ for the subset $[1,\dots,k]$ relative to the complete sample in contrast to our approach basing it only on the subset. \cite{bucher2014detecting} and \cite{kojadinovic2016testing} provide empirical evidence showing more powerful results using ranks computed relative to the subset; see also \cite{manner2019testing} for an application.       
The aforementioned papers all use averaged dependence measures as well while we allow for multiple averaged dependencies in several groups.    

\subsubsection{Binary Segmentation}\label{sec:BS}
The following break-point detection algorithm is closely related to the segmentation algorithms proposed by \cite{galeano2010shifts}, \cite{galeano2014multiple, galeano2017dating}, which, in turn, are based on the seminal work of \cite{vostrikova1981detecting}.  For a given upper tail probability \(\alpha\), the maximum of the test statistic from Eq. \eqref{eq:fluct_stat} is evaluated over \(\ell+1\) subintervals and then compared to some critical value \(c_{\alpha}(\ell)\), which depends on the amount of breaks \(\ell\) already found and which is formally derived from Proposition \ref{prop:max_lim} below. This way, the issue of multiple testing is taken into account similar to the procedure in \citet[Section 4.3]{bai1998estimating}. More formally, the algorithm can be summarized by the following two steps:

\begin{enumerate}
\item Obtain from Eq. \eqref{eq:fluct_stat} the test statistics \(\hat{M}(0,1)\). 
\begin{enumerate}
\item If the test statistic is statistically significant, i.e. if \(\hat{M}_{0,1}>c_{\alpha}(0)\), where \(c_{\alpha}(0)\) is
the asymptotic critical value \(c_{\alpha}(\ell)\) from Proposition \ref{prop:max_lim} below with \(\ell  = 0\) for a given upper tail probability \(\alpha\), then a change is announced. Let  \(z_1 = \argmmax\limits_{z \inn \Pi_T} \hat{M}(0,1;z)\) be the break point estimator and go to Step 2.
\item If the test statistic is not statistically significant, the algorithm stops, and no
change points are detected.
\end{enumerate}
\item Let \(z_1,\dots,z_\ell\) be the \(\ell\) change points in increasing order already found in previous
iterations and $z_0 \coloneqq 0$, $z_{\ell+1}\coloneqq 1$. If
\[
\mmax\limits_{1 \lleq k \lleq \ell+1}\hat{M}(z_{k-1},z_k) > c_{\alpha}(\ell),
\]
then a new change point is detected at the point fraction 
at which the value \(\hat{M}(z_{k_{\textsf{max}-1}},z_{k_{\textsf{max}}})\) is attained, where
\[
k_{\textsf{max}} = \argmmax\limits_{1 \lleq k \lleq \ell+1}\hat{M}(z_{k-1},z_k). 
\]
Repeat this step until no more change points are found.
\end{enumerate}

In other words, the whole data set is scanned for a change point, once a significant change point is found, the data set is split at that point. On the resulting subsamples the test statistic is computed again and the maximum is then compared to the according critical value. If the associated change point is significant the sample is split again at that point. This procedure stops when no significant change points can be found. This is also done in \cite{galeano2010shifts} and \cite{galeano2014multiple}.
A slightly different approach considering the subsamples seperately can be found in \cite{venkatraman1992consistency}, \cite{bai1997estimating}, \cite{aue2009break}, \cite{fryzlewicz2014wild}, among others.
Evidently, binary segmentation is commonly done in order to extend single change point detection procedures to multiple change point detection. \\

\subsubsection{Analytical Results}

In order to derive the asymptotic properties of the algorithm from Section \ref{sec:BS}, we impose the following assumptions. 

\renewcommand{\theassumption}{B}
\begin{assumption}\label{A-maxfun} For any \([a,b] \subseteq [0,1]\), the function \(M(a,b;\cdot): [a,b] \rightarrow \mathbb{R}_+\), \(z \mapsto M(a,b;z)\), where \(M(a,b;z) \coloneqq \norm{\Mb(a,b;z)}_2\), with
\[
\Mb(a,b;z) = \frac{1}{b-a}\left(\int_a^z \gb(t)\,\textnormal{\textsf{d}}t-\frac{z-a}{b-a}\int_a^b \gb(t)\,\textnormal{\textsf{d}}t\right),
\]
is either constant or has a unique maximum.
\end{assumption}

\renewcommand{\theassumption}{C}
\begin{assumption}\label{A-Lipsch} For any \(l \in \{1,\dots,H\}\), the following holds true.
\begin{enumerate}[label= \textnormal{(C\arabic*)}]
\item\label{A-Lipsch1}\(h_l: [0,1]^2 \rightarrow \mathbb{R}\) is of bounded variation in the sense of Hardy-Krause.
\item\label{A-Lipsch2}\label{A-lipschitz} \(h_l: [0,1]^2 \rightarrow \mathbb{R}\) is Lipschitz; i.e., for any \((u_1,v_1) \in [0,1]^2\) and \((u_2,v_2) \in [0,1]^2\), \(|h_l(u_1,v_1) - h_l(u_2,v_2)| \leq c_0(|u_1-u_2|+|v_1-v_2|)\) for some constant \(c_0 \in (0,\infty)\).
\end{enumerate}
\end{assumption}

Assumption \ref{A-maxfun} is similar to  \citet[Assumption 2]{galeano2014multiple} and \citet[Assumption 7]{galeano2017dating}, and restricts the nature of the breaks.
Assumption \ref{A-maxfun} is fulfilled if there is a `dominating' break in any subset of components at the same time point. For examples where the assumption is fulfilled or violated see \cite{galeano2014multiple} for the case that $\gb(t)$ is a scalar. To provide some intuition, note that, as shown in the appendix, the function \(M(a,b;z)\), specified in Assumption \ref{A-maxfun}, coincides with the probability limit of the scaled detector, i.e. \(|\hat{M}(a,b;z)/\sqrt{\floor{bT}-\floor{aT}}-M(a,b;z)| = o_p(1)\) uniformly on \([a,b]\). Assumption \ref{A-Lipsch} is satisfied by the above-mentioned dependence measures. Part \ref{A-Lipsch1} enables the use of an integration by parts formula for bivariate integrals; see, e.g. \cite{feretal2004cop} and \cite{begetal2017cop}; part \ref{A-Lipsch2} is a convenient assumption, which, however, rules out certain dependence measures like quantile dependence. 

In order to discuss the next assumption, it is convenient to first introduce the \textit{unfeasible} sequential empirical copula process 
\[
\tilde{\mathbb{B}}_k(z,\ub) \coloneqq \frac{1}{\sqrt{\floor{z_{k+1}^0T}-\floor{z_k^0T}}}\sum_{t \eq \floor{z_k^0T}}^{\floor{zT}}(1\{\Ub_t \leq \ub\}-\CC_{k+1}(\ub))
\]
for any \(z \in [z_k^0,z_{k+1}^0]\). Note that, by part \ref{A-break3} of Assumption \ref{A-break}, Theorem 2.12.1 of \cite{vw1996ep} in conjunction with the continuous mapping theorem yields \(\tilde{\mathbb{B}}_k \rightsquigarrow \mathbb{B}_k\) in \(\ell^\infty([z_k^0,z_{k+1}^0] \times [0,1]^n)\), \(k \in \{0,1,\dots,\ell^0\}\), where \(\mathbb{B}_k\) is a \(\CC_{k+1}\)-Kiefer process, i.e. \(\mathbb{B}_k\) is a tight mean-zero Gaussian process with 
\[
\cov[\mathbb{B}_k(z,\ub),\mathbb{B}_k(s,\vb)] = \varphi_k(z \wedge s)(\CC_{k+1}(\ub \wedge \vb)-\CC_{k+1}(\ub)\CC_{k+1}(\vb))
\]
and \(\varphi_k(z) \coloneqq (z-z_{k}^0)/(z_{k+1}^0-z_{k}^0)\) for any \(z,s \in [z_k^0,z_{k+1}^0]\), and \(\ub,\vb \in [0,1]^n\). The following Assumption \ref{A-dgp} ensures that the sequential \textit{residual} empirical copula process
\[
\tilde{\mathbb{C}}_k(z,\ub) \coloneqq \frac{1}{\sqrt{\floor{z_{k+1}^0T}-\floor{z_k^0T}}}\sum_{t \eq \floor{z_k^0T}}^{\floor{zT}}(1\{\hat{\Ub}_t^{\floor{z_{k}^0T}+1:\floor{z_{k+1}^0T}} \leq \ub\}-\CC_{k+1}(\ub))
\]
converges weakly in \(\ell^\infty([z_k^0,z_{k+1}^0] \times [0,1]^n)\) to 
\[
\normalfont\mathbb{C}_{k}(z,\ub) \coloneqq \mathbb{B}_k(z,\ub)-\sum_{i\eq 1}^n \mathbb{B}_k(z,\ub^{(i)}) \dot{\CC}_{i,k+1}(\ub),\;\ub^{(i)} = (1,\dots,1,u_i,1,\dots,1),
\] 
which is a crucial ingredient for the development of the asymptotic theory; see \cite{nasri2022change}. Note that \(\mathbb{C}_{k}\) is unaffected by the estimation error associated with \(\hbphi\).

\renewcommand{\theassumption}{D}
\begin{assumption}\label{A-dgp} The following holds true:
\begin{enumerate}[label= \textnormal{(D\arabic*)}]
\item\label{A-dgp1} \((\tilde{\mathbb{B}}_k,\sqrt{T}(\hbphi-\bphi^0)) \rightsquigarrow (\mathbb{B}_k,\Theta)\),  \(k \in \{0,1,\dots,\ell^0\}\), where \(\Theta\) is a tight random vector;
\item\label{A-dgp2} \(\normalfont\ssup\limits_{x \inn \mathbb{R}}|x \ff_i(x)| < \infty\), \(i \in \{1,\dots,n\}\), where \(\normalfont\ff_i = \partial\FF_i/\partial x_i\);
\item\label{A-dgp3} Eq. \textnormal{\eqref{eq: Data}} describes a stationary-ergodic finite-order AR-GARCH model;
\end{enumerate}
\end{assumption}

Assumption \ref{A-dgp} is a slightly modified set of regularity conditions that can be found in \citet[Appendix B]{nasri2022change}. Part \ref{A-dgp1} is satisfied by commonly used quasi maximum likelihood estimators of AR-GARCH models; see, e.g. \cite{fraza2004garch}; part \ref{A-dgp2} restricts the tails of the densities; part \ref{A-dgp3} could be generalized at the extent of an increase in technicalities; see \cite{nasri2022change}.

Akin to \citet[Proposition 2]{bai1997estimating} and \citet[Theorem 2]{galeano2017dating}, we first derive \(T\)-consistency of the change point estimator assuming knowledge of the number of breaks \(\ell^0\).

\begin{proposition}\label{prop:const_rate} Suppose Assumptions \textnormal{\ref{A-break}}-\textnormal{\ref{A-dgp}} hold and let \(\hat{z}_1<\dots<\hat{z}_{\ell^0}\) denote the change-points found upon executing the algorithm in Section \textnormal{\ref{sec:BS}} without testing for significance. Then, for every \(k \in \{1,2,\dots,\ell^0\}\) and \(\eps > 0\), there exists a finite constant \(M \in (0,\infty)\)  and an integer \(N \in \mathbb{N}_+\) such that for all \(T > N\),
\(\normalfont\PP(|\hat{z}_{k} - z_k^0| > M/T) < \eps.\)
\end{proposition}

\noindent Once consistency is established, Proposition \ref{prop:const_rate} can be used to derive asymptotic critical values. Similar to \citet[Section 4.3]{bai1998estimating}, Proposition \ref{prop:max_lim} summarizes the limiting distribution of the statistic under the null of \(\ell^0\) breaks.

\begin{proposition}\label{prop:max_lim} Suppose Assumptions \textnormal{\ref{A-break}}-\textnormal{\ref{A-dgp}} hold. Then,
\[\normalfont
\mmax\limits_{k \inn \{0,1,\dots,\ell^0\}}\ssup\limits_{z \inn [\hat{z}_k,\hat{z}_{k+1}]} \hat{M}(\hat{z}_{k},\hat{z}_{k+1};z) \stackrel{d}{\longrightarrow} \mmax\limits_{k \inn \{0,1,\dots,\ell^0\}}\ssup\limits_{z \inn [z_{k}^0,z_{k+1}^0]}\mathbb{M}(z_{k}^0,z_{k+1}^0;z).
\]
The limiting process \(\mathbb{M}(z_{k}^0,z_{k+1}^0;z)\), \(z \inn [z_{k}^0,z_{k+1}^0]\), is given by
\[\normalfont
\mathbb{M}(z_{k}^0,z_{k+1}^0;z) \coloneqq \sqrt{\sum_{g \eq 1}^G\sum_{l \eq 1}^H\bigg[{\displaystyle{|\mathcal{G}_g| \choose 2}}^{-1}\sum\limits_{\substack{\ijn\\i,j \inn \mathcal{G}_g}} \int_{[0,1]^2}\mathbb{D}_{k,i,j}(z,u,v)\,\dd h_l(u,v)\bigg]^2}
\]
for  \(\mathbb{D}_{k,i,j,l}(z,u_i,u_j) \coloneqq \mathbb{D}_{k}(z,\ub^{(i,j)})\), with \(\normalfont\mathbb{D}_k(z,\ub) = \mathbb{C}_{k}(z,\ub)-\varphi_k(z)\mathbb{C}_{k}(z_{k+1}^0,\ub).\)
\end{proposition}
 
\noindent Importantly, the limiting distribution is unaffected by the sampling uncertainty resulting from the first-step estimation of \(\bphi^0\). Hence, as \citet[Section 3]{nasri2022change} suggests, we can compute the critical values \(c_\alpha(\ell)\) based on the traditional \textsf{IID}-bootstrap. The use of the critical values \(c_\alpha(\ell)\) ensures a correctly sized test under the null of no further breaks.

An alternative approach for dealing with the multiple testing issue in a binary segmentation framework can be found, e.g. in \cite{galeano2010shifts} and \cite{galeano2014multiple}, using a \v{S}id\'{a}k correction on the alpha level to keep the significance level constant for multiple tests.

Finally, the number of breaks found with the help of the segmentation algorithm, \(\hat{\ell}\), say, consistently estimates the true amount \(\ell^0\) if the significance level \(\alpha\) tends to zero at a sufficiently slow rate.

\begin{proposition}\label{prop:cons_numb} Suppose Assumptions \textnormal{\ref{A-break}}-\textnormal{\ref{A-dgp}} hold. In addition, \(c_\alpha(\ell) \coloneqq c_{\alpha,T}(\ell) \rightarrow \infty\) yet \(c_\alpha(\ell) = o(\sqrt{T})\) as \(T \rightarrow \infty\). Then, \(\normalfont\PP(\hat{\ell} = \ell^0) \rightarrow 1\) as \(T \rightarrow \infty\).
\end{proposition}

Proposition \ref{prop:const_rate} shows the consistency in the location of the break while proposition \ref{prop:cons_numb} shows consistency in the number of changes. 
The assumption $c_\alpha(\ell) = o(\sqrt{T})$ is common in the literature; see \cite{bai1997estimating}, \cite{galeano2014multiple}, \cite{galeano2017dating}. By construction the algorithm induces some overestimation bounded by the alpha level. In order for that to vanish asymptotically the assumption is needed. In finite samples the statistician can choose an appropriate significance level.

\begin{remark} \label{rem: infocrit} Since bootstrapping the limiting distribution from Proposition \textnormal{\ref{prop:max_lim}} at each iteration of the algorithm is time consuming, one might resort to a computationally more efficient alternative information-criterion approach.  To sketch the procedure, fix some prespecified (large) number \(L \in \mathbb{N}_+\) that is independent of \(T\). Let \(\hat{\kb} \coloneqq (\hat{k}_1,\dots,\hat{k}_L)'\) denote (in increasing order) the break points obtained from executing the algorithm \(L+1\) iterations without checking for significance. Suppose \(L\) has been chosen large enough so that \(\ell^0 \leq L\). Then, although the number of changes has been overestimated, we can still infer consistency of a subtuple of \(\hat{\zb} = \hat{\kb}/T\) by borrowing principles from \textnormal{\cite{andrews1999consistent}}. More specifically, introduce the set of selection vectors
\(\normalfont
\mathcal{C}_{L} \coloneqq \left\{\cb = (c_1,\dots,c_{L})' \in \mathbb{R}^{L} \mid c_j \in \{0,1\},\,1 \leq j \leq L \right\}.
\)
Let \(\hat{\kb}_c  \coloneqq (\hat{k}_{1,c},\dots,\hat{k}_{|\cb|_1,c})' \in \mathbb{N}^{|\cb|_1}\) denote the \(|\cb|_1 \times 1\) vector of (in increasing order) break-point candidates selected by \(\cb \in \mathcal{C}_L\), with \(|\cb|_1 \coloneqq \sum_{j \eq 1}^{L}c_j\), where \(\cb\) is a \(L \times 1\) \textit{break-point selection vector} whose \(j\)th entry is one if a break point from \(\hat{\kb}\) is included in \(\hat{\kb}_c\) and zero otherwise. Hence, we aim at finding that selection vector \(\cb^0\) such that \(\hat{\zb}_{c^0} \stackrel{p}{\rightarrow} \zb^0\), where \(\zb^0 \coloneqq (z_{1}^0,\dots,z_{\ell^0}^0)\). To do so, we define for any \(\cb \in \mathcal{C}_L\) a break-point `information criterion'
\(\normalfont
\textsf{IC}_T(\cb) \coloneqq \hat{S}^2(\cb) + h(|\cb|_1)\kappa_T, 
\)
where \(\hat{S}(\cb) \coloneqq  \hat{S}(\kb_c)\), with
\begin{equation}\nn\normalfont
\begin{split}
\hat{S}(\kb_c)   \coloneqq \,&\mmax\limits_{1 \lleq j \lleq |\cb|_1+1}\mmax\limits_{k_{c,j-1} \,<\, t \lleq k_{c,j}}\hat{S}(k_{c,j-1},k_{c,j};t)\\
\normalfont\hat{S}(k_{c,j-1},k_{c,j};t)\coloneqq \,&\frac{t-k_{c,j-1}}{\sqrt{k_{c,j}-k_{c,j-1}}}\norm{\hat{\mb}^{k_{c,j-1}+1:t}-\hat{\mb}^{k_{c,j-1}+1:k_{c,j}}}_2,\;\;\hat{k}_{0,c} \coloneqq 1,\,\hat{k}_{|\cb|_1+1,c} \coloneqq T.
\end{split}
\end{equation}
Imposing similar conditions on \(h(\cdot)\) and the penalty \(\kappa_T \rightarrow \infty\) as in \textnormal{\cite{andrews1999consistent}} and \textnormal{\cite{luandrews2001}}, it can be shown that \((\hat{\kb}_{\hat{c}}/T,|\hat{\cb}|_1) \stackrel{p}{\longrightarrow} (\zb^0,\ell^0)\); see Appendix \textnormal{\ref{ICproof}} for details.
\end{remark}

\subsubsection{Wild Binary Segmentation}\label{sec:WBS}
A more recent approach for multiple change point detection has been proposed by \cite{fryzlewicz2014wild}, the so called \textit{wild} binary segmentation; see also  \cite{fryzlewicz2020detecting}. Since test statistics, in the context of multiple break point detection, are often tailored against single change point alternatives, these tests might not perform well in terms of power when the process contains several change points. This is due to the fact that different change points can offset each other; see, e.g. \citet[section~2]{fryzlewicz2014wild}. To counteract this issue, $m$ random intervals are considered by the wild binary segmentation algorithm. Importantly, we add also the complete subsample that would have been used by the ordinary binary segmentation of Section \ref{sec:BS} to ensure consistency. Within these intervals, the test statistic of Eq. \eqref{eq:fluct_stat} is calculated and the interval associated with the largest statistic announces a change point candidate that can be checked for significance. If a significant change point is found, the sample is split and the procedure is repeated on these subsamples similarly to the binary segmentation until no significant change point is left. In doing so, we hope that at least one `favorable' random interval designated to find a specific break is generated; i.e. an interval that contains only one change point and is as large as possible. Therefore, this procedure adapts the wild binary segmentation from \cite{fryzlewicz2020detecting} to the same type of algorithm introduced in Section \ref{sec:BS}.

\section{Monte Carlo Simulations} \label{sec: Monte Carlo Simulation}
To analyze the performance of the segmentation procedures, we conduct Monte Carlo simulations\footnote{The computations were implemented in Matlab, parallelized and performed using CHEOPS, the DFG-funded (Funding number: INST 216/512/1FUGG) High Performance Computing (HPC) system of the Regional Computing Center at the University of Cologne (RRZK)}. The setting follows closely \cite{oh2017modeling}, where a `block-equidependence' framework is proposed. More specifically, we assume that the unknown copula can be generated by an auxiliary one-factor model (see also Remark \ref{remark:factor}): 
\begin{equation}\label{MCfactor}
X_{i,t} = b_{i,t}Z_t + \e_{i,t}, \, i = 1,\dots,n,\,t=1,\dots,T.
\end{equation}
The loadings are group-specific, i.e. \(b_{i,t} = b_{j,t} = \beta_{g,t}\) if \(i,j \in \mathcal{G}_g\), \(g = 1,\dots,G\), and \(\{\mathcal{G}_1,\dots,\mathcal{G}_G\}\) partitions the cross-sectional index set. We consider \(G = 4\) groups of equal size \(|\mathcal{G}_g| = 4\), so that \(n = 16\). The common factor is distributed according to Hansen's skew \(t\)-distribution $Z_t \overset{\textsf{IID}}{\sim} Skewt(\nu, \lambda)$ with $\nu = 4$ degrees of freedom and skewness parameter $\lambda = -0.5$, while for the idiosyncratic errors $\e_{i,t} \overset{\textsf{IID}}{\sim} t(\nu)$. A similar parameterization for the common factor and the idiosyncratic errors can be found in \cite{oh2013simulated}. Throughout, three time-series dimensions are used $T\in \{$500, 1000, 1500$\}$.

In the following several change point scenarios are analyzed, where changes in the dependence measures are induced by an abrupt change in the factor loadings that drive the dependence across marginals.  Table \ref{tab: loadings} in Appendix \ref{app: Tables} contains the loadings $\beta_{g,t}$ for the according intervals. Here, group $1$ and $2$ have opposite loadings, group $3$ does not contain any change point and group $4$ contains a larger change compared to the other groups. This should reflect different behaviours of different industries.

Analyzed are change points in Spearman's $\rho$, where bivariate dependence measure are averaged among each group. This set-up leads to a vector-valued step function as in Assumption \ref{A-break1}. In a preliminary analysis (without a group structure) multiple dependence measures including Spearman's $\rho$, Spearman's footrule and Gini's $\gamma$ were analyzed. As the dependence measures are quite similar also similar results were observed. For the critical values an \textsf{IID}-bootstrap is used analogously to \citet[Section~2.4]{manner2019testing}; see also the discussion below Proposition \ref{prop:max_lim}.

Table \ref{tab: rejection freq dependencies} contains the rejection frequencies of the different algorithms. Included are the binary segmentation ($BS$), the wild binary segmentation ($WBS$) with $m=20$ random intervals and the wild binary segmentation where the complete (sub-)interval is taken as one of the `random' intervals ($WBS_{BS}$).

The empirical size of all procedures are quite similar and are close to the nominal level of $5\%$. In the $1 break$-scenario the binary segmentation outperforms the wild binary segmentations. If only one true break exists then the first step of the BS uses the test statistic \eqref{eq:fluct_stat} on the complete interval resulting in an optimal interval to detect that break, that is an interval that is as large as possible and contains a single break. 
Comparing both wild binary segmentations, $WBS_{BS}$ detects the break more often than $WBS$ as $WBS_{BS}$ contains the complete interval as well. It is noticeable that the wild binary segmentations performs worse at the beginning and the end of the sample, this might be due to fact that the random intervals are less likely to cover the ends of the sample and have a larger coverage in the middle of the sample. An increase in the amount of random intervals $m$ might improve the results at the cost of computational time, as the likelihood of a random interval that contains the break increases.

The $2 breaks$- and $3 breaks$-scenarios paint a different picture: Here, the wild binary segmentation algorithms detect more change points than the binary segmentation. Having several breaks in close proximity highlights the advantage of the former. Here, we have a positive probability that at least one random interval includes only one of the breaks and is hopefully large, which can then be used to easily detect the underlying break. Preliminary research with non-grouped data and without the large break as in $\beta_{4,t}$ shows this behaviour more clearly, leading us to believe that the large change in the group structure drives the detection. Therefore, the group structure works particularly well if there is a larger change in one group, even when there are groups with no break at all described in $\beta_{3,t}$ or groups with opposite loadings that might offset each other. 

\begin{table}
\centering
\begin{tabular}{l|c|c|c|c|c|c}
\cline{1-7}
 & size	& \multicolumn{3}{c|}{1 break}		& 2 breaks 		& 3 breaks \\
			&			& \multicolumn{1}{c}{}& \multicolumn{1}{c}{}& 				&$z_1=0.4$		& $z_1=0.3$ \\	
			&&\multicolumn{1}{c}{$z_1=0.15$}&\multicolumn{1}{c}{$z_1=0.5$}&$z_1=0.85$&$z_2=0.6$& $z_2=0.5$ $z_3=0.7$\\		
\cline{1-7}
$BS$		& 0.059		& 1& 1	 & 1				& 0.926 			& 0.999 \\
$WBS$	& 0.071 	& 0.942& 1& 0.873				&1   			& 1\\
$WBS_{BS}$ & 0.061		& 1& 1 & 0.983 & 1 & 1
\end{tabular}
\captionof{table}{Rejection frequencies of a DGP with 0,1,2 and 3 breaks of 1000 repetitions with $T=1000$ using Spearman's $\rho$}
\label{tab: rejection freq dependencies}
\end{table}

\begin{table}[htbp]
\centering
\begin{tabular}{|c|c|ccc|c|c|}
\cline{1-7}
\multirow{3}{*}{$BS$}		&0 breaks & \multicolumn{3}{c|}{1 break}& 2 breaks& 3 breaks \\
	     	&& $z_1$=0.15& $z_1$=0.5& $z_1$=0.85	 & $z_1$=0.4 & $z_1$=0.3 \\
	     	&&&&&$z_2$=0.6&$z_2$=0.5 $z_3$=0.7 \\
\cline{1-7}
-3		&	  &		&		& & & 0.001	\\
-2		&	  &		&		&& 0.074	& 0.000\\
-1		&	  & 0.000 & 0.000 & 0.000 & 0.000	&0.047\\
0		&0.941& 0.588 & 0.959 & 0.822 & 0.841&0.917\\
1		&0.054& 0.397 & 0.040 & 0.173 & 0.082&0.034\\
2		&0.005& 0.015 & 0.001 & 0.005 & 0.003&0.001\\
3		&0.000& 0.000 & 0.000 & 0.000 & 0.000&0.000\\
\cline{1-7}
\multicolumn{1}{|c}{\multirow{2}{*}{$WBS$}}		&\multicolumn{1}{c}{}& \multicolumn{3}{c}{}			 &\multicolumn{1}{c}{}&\\\multicolumn{1}{|c}{}&\multicolumn{1}{c}{}&&&\multicolumn{1}{c}{}&\multicolumn{1}{c}{}&\multicolumn{1}{c|}{}\\			
\cline{1-7}
-3		&	  &		&		&&  &0.000\\
-2		&	  &		&		&&	0.000 & 0.096\\
-1		&	  & 0.058 & 0.000 & 0.127 & 0.016 &0.468	\\
0		&0.928& 0.840 & 0.959 & 0.823 & 0.931&0.418\\
1 		&0.068& 0.100 & 0.041 & 0.048 & 0.049&0.017\\
2		&0.003& 0.002 & 0.000 & 0.002 & 0.004&0.001\\
3		&0.001& 0.000 & 0.000 & 0.000 & 0.000 &0.000\\
\cline{1-7}
\multicolumn{1}{|c}{\multirow{2}{*}{$WBS_{BS}$}}		&\multicolumn{1}{c}{}& \multicolumn{3}{c}{}			&\multicolumn{1}{c}{}& \\ \multicolumn{1}{|c}{}&\multicolumn{1}{c}{}&&&\multicolumn{1}{c}{}&\multicolumn{1}{c}{}&\multicolumn{1}{c|}{}\\		
\cline{1-7}
-3		&	  &		&		&	&	    &0.000\\
-2		&	  &		&		&& 0.000	&0.015\\
-1		&	  & 0.000 & 0.000 & 0.017 & 0.001 &0.346\\
0		&0.939& 0.787 & 0.962 & 0.911 & 0.944 &0.609\\
1 		&0.058& 0.210 & 0.035 & 0.069 & 0.053 &0.029\\
2		&0.002& 0.003 & 0.003 & 0.003 & 0.002 &0.001\\
3		&0.001& 0.000 & 0.000 & 0.000 & 0.000 &0.000\\
\cline{1-7}
\end{tabular}
\captionof{table}{Over- and underestimation frequencies of the amount of true change points $\hat{\ell}-\ell^0$ with $\alpha = 0.05$ and $T=1000$}
\label{tab: over-/underestimation dependencies T=1000 alpha=0.05}
\end{table}

For a more in depth analysis of the estimated amount of change points, Table \ref{tab: over-/underestimation dependencies T=1000 alpha=0.05} shows the frequency of under- or overestimation of underlying breaks for the four scenarios. 
Although in the $1 break$-scenario the binary segmentation has a higher detection rate, more overestimation in the amount of changes can be found when the breaks are close to the beginning or end of the sample, showing that in this case the wild binary segmentation performs better even for the $1 break$-scenario; with $WBS_{BS}$ performing slightly better than $WBS$.

Turning to the $2 breaks$-scenario, $WBS$ detects at least one change point, while $BS$ in some cases does not detect any breaks. This result becomes more apparent in small sample sizes (compare Table \ref{tab: over-/underestimation dependencies T=500 alpha=0.1} in Appendix \ref{app: Tables}). Again, $WBS_{BS}$ improves the performance of $WBS$. The improvement arises since $WBS_{BS}$ is able to detect the remaining breakpoint after the first one is found. This is due to the fact, that after the first break is found, there is only a single break left for which one of the complete subinterval is tailor-made for it's detection. Looking at the $BS$, we see that if breaks are found, then both of them. Same argumentation as for the $WBS_{BS}$ case applies. However, quite often no break is found in small samples (or in cases with less prominent breaks), showing that several breaks in close proximity can offset each other leading to no detection. This issue is further explained and illustrated in \cite{fryzlewicz2014wild}. A clear disadvantage of the widely used binary segmentation. At the same time, this shows the strength of the wild binary segmentation.

Turning to the $3breaks$-scenario, $BS$ correctly detects the amount of breaks more often than the wild binary segmentation algorithms. In previous simulations with no groups and less prominent breaks, $BS$ found the correct amount more often than the wild binary segmentation algorithms as well but also the case that no changes at all were detected appeared more often. The same can be hinted at in Table \ref{tab: over-/underestimation dependencies T=500 alpha=0.1} in Appendix \ref{app: Tables}. Under closer inspection, $BS$ usually first finds (if the test statistic is rejected) the middle break $z_2$; see change in Spearman's $\rho$ in Table \ref{tab: loadings}. The resulting subintervals contain only one breakpoint each, which the algorithm can easily detect, explaining the results.

\begin{table}[tbp]
\centering
\begin{tabular}{|l|ccc|c|c|}
\cline{1-6}
\multirow{3}{*}{$BS$}		& \multicolumn{3}{c|}{1 break}& 2 breaks& 3 breaks \\
	     	& $z_1$=0.15& $z_1$=0.5& $z_1$=0.85	 & $z_1$=0.4 & $z_1$=0.3 \\
	     	&&&&$z_2$=0.6&$z_2$=0.5 $z_3$=0.7 \\
\cline{1-6}
$T=500$ & 9.43 & 2.38 & 6.96 & 2.58 & 8.91\\
$T=1000$& 6.11 & 1.38 & 3.66 & 2.03 & 3.09\\
$T=1500$& 5.33 & 0.82 & 2.55 & 1.32 & 1.46\\
\cline{1-6}
\multicolumn{1}{|c}{\multirow{2}{*}{$WBS$}}		& \multicolumn{3}{c}{}			 &\multicolumn{1}{c}{}&\\\multicolumn{1}{|c}{}&&&\multicolumn{1}{c}{}&\multicolumn{1}{c}{}&\multicolumn{1}{c|}{}\\			
\cline{1-6}
$T=500$ & 5.33 & 2.52 & 4.53 & 7.78 & 18.86\\
$T=1000$& 3.62 & 1.47 & 1.95 & 1.99 & 12.29\\
$T=1500$& 3.15 & 0.79 & 1.67 & 0.95 & 4.34\\
\cline{1-6}
\multicolumn{1}{|c}{\multirow{2}{*}{$WBS_{BS}$}}		& \multicolumn{3}{c}{}			&\multicolumn{1}{c}{}& \\ \multicolumn{1}{|c}{}&&&\multicolumn{1}{c}{}&\multicolumn{1}{c}{}&\multicolumn{1}{c|}{}\\		
\cline{1-6}
$T=500$ & 8.34 & 2.62 & 5.44 & 4.68 & 17.66\\
$T=1000$& 5.45 & 1.41 & 2.89 & 1.90 & 8.52\\
$T=1500$& 4.73 & 0.76 & 1.87 & 0.83 & 1.94\\
\cline{1-6}
\end{tabular}
\captionof{table}{Hausdorff distance $d_H$  for $T=1000$ averaged over 1000 iterations multiplied by 100 }
\label{tab: Hausdorff}
\end{table}

Following \cite{wang2020statistical} or \cite{okui2021heterogeneous}, among others, Table \ref{tab: Hausdorff} shows the accuracy of the location of the estimated breaks measured by the average Hausdorff distance $d_H$ of the change point fractions (multiplied by $100$):
\begin{align*}
d_H(\bm{z}^0, \hat{\bm{z}}) = \max \left\lbrace \max_{1 \leq j \leq \ell^0} \min_{1 \leq k \leq \hat{\ell}}|z^0_j - \hat{z}_k| , \max_{1 \leq k \leq \hat{\ell}} \min_{1 \leq j \leq \ell^0} |z^0_j - \hat{z}_k| \right\rbrace
\end{align*}
with $\bm{z}^0 \coloneqq \{z_1^0,\dots,z_{\ell^0}^0\}$ and $\hat{\bm{z}} \coloneqq \{\hat{z}_1,\dots,\hat{z}_{\hat{\ell}}\}$. The accuracy measured by the Hausdorff distance confirms our previous findings. Moreover, Table \ref{tab: Hausdorff} provides finite sample evidence for the consistency of the segmentation algorithms as the Hausdorff distance converges to zero with sample size.   

Finally, for the empirical application of the following section, residuals of the DGP in (\ref{eq: Data}) are used by estimating AR-GARCH parameters by quasi maximum likelihood. In order to show that the estimation error from this filtration does not distort the results, we follow \cite{oh2013simulated} and consider an AR(1)-GARCH(1,1) structure with Gaussian innovations $\eta_{i,t}$:
\begin{align*}
Y_{i,t} = \phi_0 + \phi_1 Y_{i,t-1} + \sigma_{it}\eta_{i,t},\quad \text{ with }\quad\sigma_{i,t}^2 = \omega + \beta \sigma_{i,t-1}^2 + \alpha \sigma_{i,t-1}^2\eta_{i,t-1}^2,
\end{align*}
where  $\phi = (\phi_0,\phi_1,\omega,\beta, \alpha)' = (0.01,0.05,0.05,0.85,0.1)'$. The cross-sectional dependence structure of the innovations \(\bm{\eta}_t \coloneqq (\eta_{1,t},\dots,\eta_{n,t})'\), is governed by the factor copula \eqref{MCfactor}.
Table \ref{tab: rejection freq dependencies, filtered} summarizes rejection frequencies using filtered data. Table \ref{tab: over-/underestimation dependencies T=1000 alpha=0.05, filtered} in Appendix \ref{app: Tables} shows the over- and underestimation of the change points.        
\begin{table}
\centering
\begin{tabular}{l|c|c|c|c|c|c}
\cline{1-7}
 & size	& \multicolumn{3}{c|}{1 break}		& 2 breaks 		& 3 breaks \\
			&			& \multicolumn{1}{c}{}& \multicolumn{1}{c}{}& 				&$t_1=400$		& $t_1=300$ \\	
			&&\multicolumn{1}{c}{$t_1=150$}&\multicolumn{1}{c}{$t_1=500$}&$t_1=850$&$t_2=600$& $t_2=500$ $t_3=700$\\		
\cline{1-7}
$BS$		& 0.056		& 1& 1	 & 1				& 0.965 			& 1 \\
$WBS$	& 0.050 	& 0.940& 1& 0.861				&1   			& 1\\
$WBS_{BS}$ & 0.050		& 1& 1 & 0.981 & 1 & 1
\end{tabular}
\captionof{table}{Rejection frequencies of filtered data with 0,1,2 and 3 breaks of 1000 repetitions with $T=1000$ using Spearman's $\rho$}
\label{tab: rejection freq dependencies, filtered}
\end{table}
As can be seen, the results are quite similar to the case of no filtration.

\section{Empirical Application} \label{sec: Empirical Application}
For our empirical application, data from Thomson Reuters Eikon is used that covers the recent COVID-19 pandemic where financial markets plummeted and quickly recovered, giving us reason to believe in at least one change point of the dependence structure between assets due to a diversification meltdown. More specifically, return data of companies from the four largest industry sectors of the EURO STOXX 50 is used from $01.01.2016$ until $30.06.2021$ resulting in $T= $ 1,433 trading days and $n=31$ assets; see Table \ref{tab: emp.appl.-companies}. Any serial dependence and GARCH effects of the marginals are filtered out using an AR(1)-GARCH(1,1) specification with $t$-innovation\footnote{Averaging over each of the 31 estimated parameters results in a process of the form $Y_t = 0.0002-0.0377Y_{t-1}+\e_t,$ with $\sigma^2_t = 0.0000+0.8830\sigma^2_{t-1} + 0.0891\e^2_{t-1}$.}.

\begin{table}
\centering
\begin{tabular}{|l l|}
\cline{1-2}
\multirow{4}{*}{Finance} & Allianz, Axa, Banco Santander, BBVA, Deutsche Bank, \\
&Deutsche Boerse, BNP Paribas, Generali, ING Groep, \\
&Intesa Sanpaolo, Muenchner Rueck., Societe Generale, \\
&Unicredit  \\
\cline{1-2}
\multirow{2}{*}{Energy} & E.ON, Enel, Eni, Suez, Iberdrola, Repsol, RWE, \\
&TotalEnergies\\
\cline{1-2}
\multirow{2}{*}{Telecom and Media} & Deutsche Telekom, Orange, Telecom Italia, Telefonica, \\
& Vivendi\\
\cline{1-2}
Consumer Retail & Anheuser-Busch, Carrefour, Danone, L'Oreal, LVMH\\
\cline{1-2}
\end{tabular}
\captionof{table}{Included Stocks by industry}
\label{tab: emp.appl.-companies}
\end{table}  

A significance level of $\alpha=0.05$ is chosen with $B=500$ bootstrap repetitions. For the WBS algorithm, $50$ random intervals are used.
Table \ref{tab: emp. appl.-breakpoints} shows the change points found for all three procedures.
\begin{table}[tbp]
\centering
\begin{tabular}{c|c|c}
$BS$ 	& $WBS$ 	& $WBS_{BS}$ \\
\cline{1-3}
7.11.16	& 7.11.16	& 7.11.16	\\
		& 23.01.20	& 6.04.18	\\
		& 21.09.20	& 20.02.20	\\
		&			& 21.09.20	\\		
\end{tabular}
\captionof{table}{Breakpoints}
\label{tab: emp. appl.-breakpoints}
\end{table}
\begin{figure}[t]
  \begin{subfigure}[b]{0.5\textwidth}
	\includegraphics[width=\textwidth]{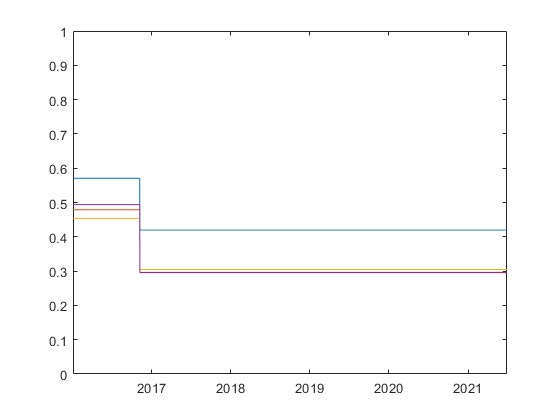}
    \caption{$BS$}
    \label{subfig: Avg_Rank_Corr-BS}
  \end{subfigure}
  \hfill
  \begin{subfigure}[b]{0.5\textwidth}
	\includegraphics[width=\textwidth]{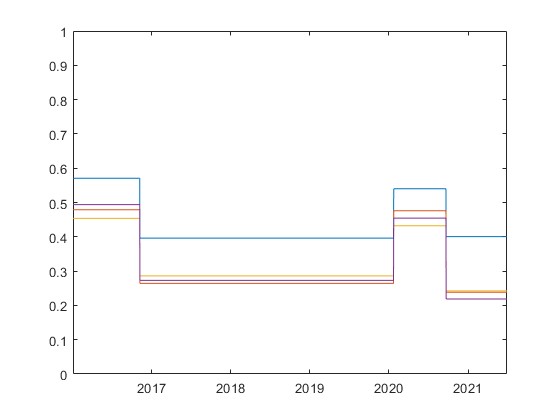}
    \caption{$WBS$}
    \label{subfig: Avg_Rank_Corr-WBS}
  \end{subfigure}
  \hfill
  \begin{subfigure}[b]{\textwidth}
  \centering
	\includegraphics[width=0.5\textwidth]{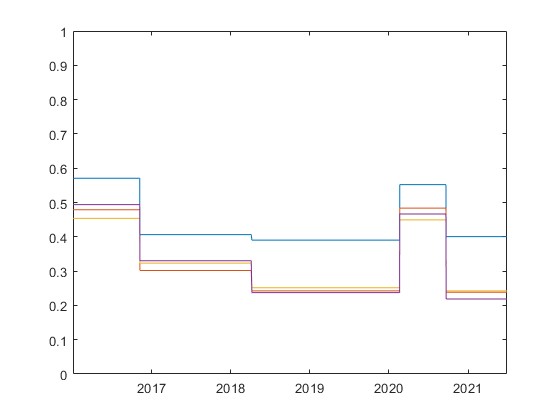}
    \caption{$WBS_{BS}$}
    \label{subfig: Avg_Rank_Corr-WBSBS}
  \end{subfigure}
  \caption{Rank correlation averaged among industries}
  \centering
  \begin{tabular}{r@{: }l r@{: }l}
	blue& Finance & red & Energy\\
	purple& Consumer Retail & yellow& Telecom and Media 
\end{tabular}
  \label{fig: Avg_Rank_Corr}
\end{figure}
All procedures find a change point at 07.11.2016 which closely coincides with the presidential election 2016 and the election of Donald Trump as 45th US President on the 8th of November 2016. The stock market not only in the US but also in Europe reacted bullish after the election. The WBS algorithms additionally find breaks at the beginning of the coronavirus pandemic in accordance with the lockdown of Wuhan in January 2020, while the break date 20.02.2020 coincides with a massive price drop of the EURO STOXX 50. The different detection time of both algorithms might be explained by the use of different random intervals. Note, that the second change point candidate of the $BS$ is the same as the one from $WBS$ (23.01.2020), with a $p$-value of $0.072$.   Both $WBS$ algorithms also detect a break at 21.09.2020 which coincides with the second wave of the COVID-19 pandemic and the resulting implementation of new restrictions; e.g. 20.09.2020 and 14.10.2020 new restrictions were announced in the UK and France, respectively. The $WBS_{BS}$ algorithm additionally finds a break at 06.04.2018, which corresponds with a sell-off starting on Wall Street and spreading to Europe resulting in a price drop in assets at the beginning of February 2018.

The WBS algorithms find quite similar change points, while the BS algorithm only finds a single change point. To provide some intuition: After $BS$ finds the first change point at the beginning of the data set, a situation with two remaining changes (in the second subsample) emerge for which the binary segmentation either finds both changes or misses both, which is in line with the results of the Monte Carlo simulations for two underlying change points. Figure \ref{subfig: Avg_Rank_Corr-WBS}, \ref{subfig: Avg_Rank_Corr-WBSBS} show the estimated dependence measures between the changes revealing a dependence measure process similar to the dgp of Section \ref{sec: Monte Carlo Simulation}. Finally, the $p$-values of $WBS$ and $WBS_{BS}$ after the change points have been found are $0.236$ and $0.342$, respectively, suggesting we do not miss any further change point.

\begin{figure}[!tbp]
  \begin{subfigure}[b]{0.5\textwidth}
	\includegraphics[width=\textwidth]{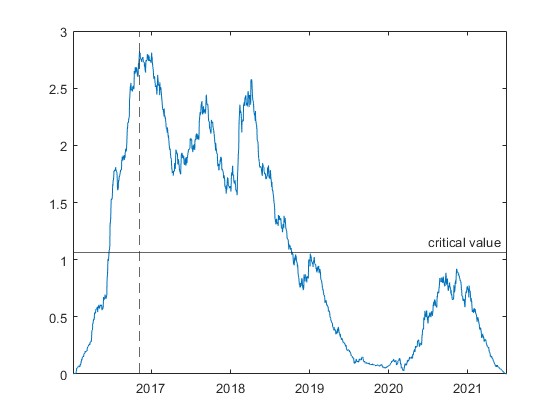}
    \caption{$BS$}
    \label{subfig: 2Breaks-BS}
  \end{subfigure}
  \hfill
  \begin{subfigure}[b]{0.5\textwidth}
	\includegraphics[width=\textwidth]{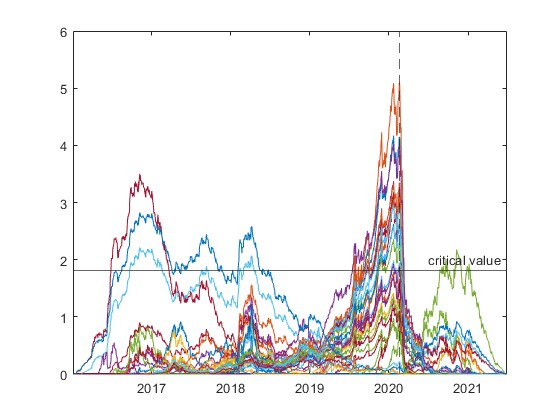}
    \caption{$WBS_{BS}$}
    \label{subfig:2Breaks-WBSBS}
  \end{subfigure}
  \caption{Test statistic for all time points using (a) binary segmentation and (b) wild binary segmentation with $50$ random intervals. The dashed line marks the maximal value of the test statistic and therefore the first change point candidate.}
  \label{fig: test statistic}
\end{figure}

Figure \ref{fig: test statistic} illustrates the test statistic of the binary ($BS$) and wild binary segmentation ($WBS_{BS}$) in the first step to find the first change point. $BS$ finds the first break at the end of 2016. Although the test statistic shows signs of a change at the end of 2020, it does not get very large. The test statistic for $WBS_{BS}$ announces a first change point at the beginning of 2020. However, also for the other change points the test statistic exceeds the critical value. Notice that $WBS_BS$ includes the test statistic of the binary segmentation.

\section{Conclusion} \label{sec: Conclusion}
Procedures for the detection of multiple change points for copula-based dependence measures have been analyzed. Focusing on the binary segmentation, consistency results for the amount as well as the location of change points are provided. Additionally, a consistency result for a computationally more efficient information criterion is given. In order to show that the procedures perform well in finite samples, a Monte Carlo simulation is conducted. To generate data, a factor copula with piecewise constant, group specific factor loadings is used. A binary segmentation and two wild binary segmentation algorithms are implemented in order to detect changes in Spearman's $\rho$.  The proposed procedures are then applied on daily return data of companies, included in the EURO STOXX 50 index, to study change points in Spearman's $\rho$ covering the recent COVID-19 pandemic. The wild binary segmentation algorithms find change points in the dependencies, among others, at the beginning as well as at the second wave of the COVID-19 pandemic.

One interesting avenue for future research could be a more detailed investigation of the information criterion of  Remark \ref{rem: infocrit} to find a suitable penalty term leading to a more time efficient procedure for finding multiple change points. The consistency results in this article might provide the foundation of this method. Moreover, checking for breaks in the marginals [using, e.g. \cite{galeano2010shifts} or \cite{wied2012new}] in conjunction with a segmentation algorithm to estimate the marginals between the detected changes, could be another interesting question.

\clearpage  
\appendix

\begin{appendices}

\section[Appendix A: Proofs]{Proofs}
\setcounter{equation}{0}
\numberwithin{equation}{section}

\subsection{Proof of Proposition \ref{prop:const_rate}}
\textbf{Proof.} Proposition \ref{prop:const_rate} follows from Lemma \ref{lem-aux1} and \ref{lem-aux2} below, in conjunction with the proof of \citet[Theorem 1]{galeano2017dating}.

\renewcommand{\thelemma}{A.1}
\begin{lemma}\label{lem-aux1}
For any given interval \([a,b] \subseteq [0,1]\),
\[
\textnormal{$\ssup\limits_{z \inn [a,b]}$}\left\vert\frac{\hat{M}(a,b;z)}{\sqrt{\floor{bT}-\floor{aT}}}-M(a,b;z)\right\vert = O_p(1/\sqrt{T}).
\]
\end{lemma}

\noindent\textbf{Proof of Lemma \ref{lem-aux1}.} Consider
\[
\frac{\hat{M}(a,b;z)}{\sqrt{\floor{bT}-\floor{aT}}} = \sqrt{\sum_{g \eq 1}^G\sum_{l \eq 1}^H\left[\frac{1}{\displaystyle{|\mathcal{G}_g| \choose 2}}\sum\limits_{\substack{\ijn\\i,j \inn \mathcal{G}_g}}\frac{\floor{zT}-\floor{aT}}{\floor{bT}-\floor{aT}}(\hat{m}_{i,j,l}^{\floor{zT}+1:\floor{bT}}-\hat{m}_{i,j,l}^{\floor{aT}+1:\floor{bT}})\right]^2}, 
\]
fix an arbitrary pair \((i,j) \in \mathcal{G}_g, \ijn\), for some \(g \in \{1, \dots, G\}\). Next, choose some \(l \in \{1,\dots,H\}\) and define for any given interval \([a,b] \subseteq [0,1]\)  the \textit{unfeasible} dependence measure
\begin{equation}
\tilde{m}_{i,j,l}^{\floor{a T}+1:\floor{b T}} = \frac{1}{\floor{b T}-\floor{a T}}\sum_{t \eq \floor{a T}+1}^{\floor{b T}}h_l(U_{i,t},U_{j,t}).
\end{equation}
Note that, by the triangle inequality in conjunction with Assumptions \ref{A-Lipsch}, one gets
\begin{equation}\label{eq:mtilde}
\begin{split}
|\hat{m}_{i,j,l}^{\floor{a T}+1:\floor{b T}} - \tilde{m}_{i,j,l}^{\floor{a T}+1:\floor{b T}}|	 \leq &\,\frac{c_0}{\floor{b T}-\floor{a T}}\sum_{t \eq \floor{a T}+1}^{\floor{b T}}\sum_{k \inn \{i,j\}} |\hat{U}_{k,t}^{\floor{aT}+1:\floor{bT}}-U_{k,t}|  \\
= &\, \frac{c_0}{\floor{b T}-\floor{a T}}\sum_{t \eq \floor{a T}+1}^{\floor{b T}}\frac{1}{\sqrt{T}}\sum_{k \inn \{i,j\}}|\hat{\mathbb{F}}_{k}(a,b,\eta_{k,t})|
\end{split}
\end{equation}
where 
\[
\hat{\mathbb{F}}_i(a,b,x_i) = \hat{\mathbb{F}}(a,b,\xb^i),\;\; \xb^i = (\infty,\dots,\infty,x_i,\infty,\dots,\infty)
\]
for \(i \in \{1,\dots,n\}\), with \(\hat{\mathbb{F}}(a,b,\xb) = \sqrt{T}(\hat{F}^{\floor{aT}+1:\floor{bT}}-\textnormal{\textsf{F}})(\xb)\) for \(\xb \in \mathbb{R}^n\) and  \((a,b) \in \Delta\), with \(\Delta = \Delta_{0,1}\), where \(\Delta_{a,b} = \{(\alpha,\beta) \in [a,b]^2: 0 \leq a \leq b \leq 1\}\) and
\[
\hat{F}^{\floor{aT}+1:\floor{bT}}(\xb) = \frac{1}{\floor{bT}-\floor{aT}}\sum_{t = \floor{a T} +1}^{\floor{b T}}1\{\hat{\etab}_{t} \leq \xb\}.
\]
Now, by Assumption \ref{A-break}, \(\mmax_{k \inn \{i,j\}}\ssup_{(a,b,x) \inn \Delta \times \mathbb{R}}\,|\hat{\mathbb{F}}_{k}(a,b,x)| = O_p(1)\). Hence, by Eq. \eqref{eq:mtilde}, \(\ssup_{(a,b) \inn \Delta}|\hat{m}_{i,j}^{\floor{a T}+1:\floor{b T}}- \tilde{m}_{i,j}^{\floor{a T}+1:\floor{b T}}|= O_p(1/\sqrt{T}).\) Next, assume, for brevity\footnote{The following can be readily extended to \(\ell^0 > 2\): Similar to the case with \(\ell^0 = 2\) break points, one has to distinguish between \(\ell^0+1+\ell^0(\ell^0+1)/2\) cases; i.e., \(\ell^0 +1\) cases with no breaks located within \([a,z]\) and \(\ell^0(\ell^0+1)/2\) cases with at least one break located within \([a,z]\).}, two \(\ell^0 = 2\) break points \(0 \eqqcolon z_0^0 < z_1^0 < z_2^0 < z_3^0 \coloneqq 1\), \(n = 2\), and \(H = 1\); i.e., \(p = 1\). Then either \textbf{1.} \(a \leq z_1^0 < z_2^0 \leq b\), \textbf{2.} \(z_2^0 \leq a\), \textbf{3.} \(z_1^0 \leq a < z_2^0 \leq b\), \textbf{4.} \(z_1^0 \leq a < b \leq z_2^0\), \textbf{5.} \(a \leq z_1^0 < b \leq z_2^0\), or \textbf{6.} \(z_1^0 \geq b\).  Hence,
\begin{equation}
\begin{split}
\int_a^zg(t)\,\textsf{d}t = \,&  1\{a \leq z_1^0 < z_2^0 \leq z\}[(z_1^0-a)\gamma_{0}^0 + (z_2^0-z_1^0)\gamma_{1}^0 + (z-z_2^0)\gamma_{2}^0] \\
\,& + 1\{a \leq z_1^0 < z \leq z_2^0\}[(z_1^0-a)\gamma_{0}^0+(z-z_1^0)\gamma_{1}^0]\\
\,& + 1\{z_1^0 \leq a < z_2^0 \leq z\}[(z_2^0-a)\gamma_{1}^0+(z-z_2^0)\gamma_{2}^0]\\
\,& + (b-a)[1\{z_1^0 \geq z\}\gamma_{0}^0+ 1\{z_1^0 \leq a < z \leq z_2^0\}\gamma_{1}^0+ 1\{z_2^0 \leq a\}\gamma_{2}^0]. 
\end{split}
\end{equation}
Now, consider
\begin{equation}
\begin{split}
(\floor{z T}-\floor{a T})&\tilde{m}_{i,j}^{\floor{aT}+1:\floor{zT}} \\
 =  \,& 1\{a \leq z_1^0 < z_2^0 \leq z\}\bigg[(\floor{z_1^0 T}-\floor{a T})\tilde{m}_{i,j}^{\floor{aT}+1:\floor{z_1^0T}} \\
\,& \hspace*{3.5cm} + (\floor{z_2^0 T}-\floor{z_1^0 T})\tilde{m}_{i,j}^{\floor{z_1^0T}+1:\floor{z_2^0T}}\\
\,& \hspace*{3.5cm} +  (\floor{z T}-\floor{z_2^0 T})\tilde{m}_{i,j}^{\floor{z_2^0T}+1:\floor{zT}}\bigg] \\
\,& + 1\{a \leq z_1^0 < z \leq z_2^0\}\bigg[ (\floor{z_1^0 T}-\floor{a T})\tilde{m}_{i,j}^{\floor{aT}+1:\floor{z_1^0T}} \\
\,& \hspace*{3.5cm} + (\floor{z T}-\floor{z_1^0 T})\tilde{m}_{i,j}^{\floor{z_1^0T}+1:\floor{zT}}\bigg] \\
\,& + 1\{z_1^0 \leq a < z_2^0 \leq z\}\bigg[(\floor{z_2^0 T}-\floor{a T})\tilde{m}_{i,j}^{\floor{aT}+1:\floor{z_2^0T}} \\
\,& \hspace*{3.5cm} + (\floor{z T}-\floor{z_2^0 T})\tilde{m}_{i,j}^{\floor{z_2^0T}+1:\floor{zT}}\bigg] \\
\,& + (1\{z_1^0 \geq z\}+ 1\{z_1^0 \leq a < z \leq z_2^0\}+1\{z_2^0 \leq a\})\tilde{m}_{i,j}^{\floor{aT}+1:\floor{zT}}. 
\end{split}
\end{equation}
Next, introduce the \textit{unfeasible} sequential empirical copula process 
\[
\tilde{\mathbb{B}}_{i,j,t}(a,b,u,v) = \frac{1}{\sqrt{\floor{bT}-\floor{aT}}}\sum_{t \eq \floor{aT}}^{\floor{bT}}(1\{U_{i,t} \leq u, U_{j,t} \leq v\}-\CC_{i,j,t}(u,v)).
\] 
By part \ref{A-break1} of Assumption \ref{A-break}, one has \(\textsf{C}_{i,j,t} = \textsf{C}_{i,j,k}\) for all \(\floor{aT} \leq t \leq \floor{bT}\) and thus
\(\tilde{\mathbb{B}}_{i,j,t}(\alpha,\beta,u,v) = \tilde{\mathbb{B}}_{i,j,k}(\alpha,\beta,u,v)\) for any \((\alpha,\beta) \in \Delta_{a,b}\), \([a,b] \subseteq [z_{k}^0,z_{k+1}^0]\), \(k \in \{0,1,2\}\), and 
\(
m_{i,j,t} = \gamma_k^0 = \int_{[0,1]^2}h(u,v) \,\textsf{d}\textsf{C}_{i,j,k}(u,v).
\)
Moreover, by \citet[Theorem 6]{feretal2004cop}, one gets
\[
\sqrt{T}(\tilde{m}_{i,j}^{\floor{\alpha T}+1:\floor{\beta T}}-m_{i,j,k})  = \int_{[0,1]^2}\tilde{\mathbb{B}}_{i,j,k}(\alpha,\beta,u,v)\, \textsf{d}h(u,v)+O(1/\sqrt{T}).
\]
Since, by Assumption \ref{A-dgp}, \(\tilde{\mathbb{C}}_{i,j,k}\) converges weakly for any \(k \in \{1,\dots,\ell\}\) on \(\Delta_{a,b} \times [0,1]^2\) to a tight Gaussian process [see also \cite{bucherko2016cop}], the claim follows. \hfill \qed

\renewcommand{\thelemma}{A.2}
\begin{lemma}\label{lem-aux2} Define \(\normalfont\hat{z}_{a,b} = \argmmax\limits_{z \in [a,b]} \hat{M}(a,b;z)\). For any \([a,b] \subseteq [0,1]\), the change point estimator \(\hat{z}_{a,b}\)  is consistent for the dominating change point in \([a,b]\) defined as \(\normalfont z_{a,b}^\star \coloneqq \argmmax\limits_{z \inn [a,b]} M(a,b;z)\).
\end{lemma}

\subsection{Proof of Proposition \ref{prop:max_lim}}
\noindent \textbf{Proof.} The proof follows from part Lemma \ref{lem-auxB1} and part \ref{lema-auxB21} of Lemma \ref{lem-auxB2} below.

\renewcommand{\thelemma}{B.1} 
\begin{lemma}\label{lem-auxB1} If there is no break in \([z_k^0,z_{k+1}^0]\), \(k \in \{0,1,\dots,\ell^0\}\), then
\[\normalfont
\mmax\limits_{k \inn \{0,1,\dots,\ell^0\}}\ssup\limits_{z \inn [z_{k}^0,z_{k+1}^0]} \hat{M}_{T}(z_{k}^0,z_{k+1}^0;z) \stackrel{d}{\longrightarrow} \mmax\limits_{k \inn \{0,1,\dots,\ell^0\}}\ssup\limits_{z \inn [z_{k}^0,z_{k+1}^0]}\mathbb{M}(z_{k}^0,z_{k+1}^0;z).
\]
\end{lemma}

\noindent \textbf{Proof of Lemma \ref{lem-auxB1}.} Set \(\hat{M}_{k}(z) \coloneqq \hat{M}(z_k^0,z_{k+1}^0;z)\). Now, for some \(k \in \{0,1,\dots,\ell^0\}\), \(l \in \{1,\dots,H\}\), \(g \in \{1,\dots,G\}\), and any pair \((i,j) \in \mathcal{G}_g\), \(i \neq j\), one has
\begin{equation}
\begin{split}
\hat{m}_{i,j,l}^{\floor{z_k^0T}+1:\floor{zT}}-&\hat{m}_{i,j,l}^{\floor{z_k^0T}+1:\floor{z_{k+1}^0T}} \\
 \,&= \int_{[0,1]^2}(\hat{C}_{i,j}^{\floor{z_k^0 T}+1:\floor{zT}}-\hat{C}_{i,j}^{\floor{z_k^0 T}+1:\floor{z_{k+1}^0T}})(u,v)\,\dd h_l(u,v) + O_p(1/T),
\end{split}
\end{equation}
using \cite{feretal2004cop}; see also \cite{begetal2017cop}. Moreover, consider
\begin{equation}
\begin{split}
\hat{\mathbb{D}}_{k}(z,\ub) \coloneqq \,& \frac{\floor{zT}-\floor{z_k^0T}}{\sqrt{\floor{z_{k+1}^0T}-\floor{z_k^0T}}}(\hat{C}^{\floor{z_k^0 T}+1:\floor{zT}}-\hat{C}^{\floor{z_k^0 T}+1:\floor{z_{k+1}^0T}})(\bm{u})\\ = \,& \hat{\mathbb{C}}_{k}(z,\ub) - \varphi_k(z)\hat{\mathbb{C}}_{k}(z_{k+1}^0,\ub),
\end{split}
\end{equation}
where 
\begin{equation}
\hat{\mathbb{C}}_{k}(z,\ub) \coloneqq \frac{1}{\sqrt{\floor{z_{k+1}^0T}-\floor{z_k^0T}}}\sum_{t \eq \floor{z_k^0T}+1}^{\floor{zT}}(1\{\hat{\bm{U}}_{t}^{\floor{z_k^0 T}+1:\floor{zT}} \leq \bm{u}\} - \CC_{k+1}(\ub))
\end{equation}
for \(z \in [z_k^0,z_{k+1}^0]\). We can deduce from \cite{bucherko2016cop} and \cite{nasri2022change} in conjunction with the continuous mapping theorem, that \(\hat{\mathbb{C}}_{k} \rightsquigarrow  \mathbb{C}_{k}\) in \(\ell^\infty([z_k^0,z_{k+1}^0] \times [0,1]^n)\). Hence, \(\hat{\mathbb{D}}_{k} \rightsquigarrow \mathbb{D}_k\) in \(\ell^\infty([z_k^0,z_{k+1}^0] \times [0,1]^n)\), where \(\mathbb{D}_k(z,\ub) = \mathbb{C}_{k}(z,\ub)-\varphi_k(z)\mathbb{C}_{k}(z_{k+1},\ub).\) Therefore, 
\begin{equation}
\begin{split}
\frac{\floor{zT}-\floor{z_k^0T}}{\sqrt{\floor{z_{k+1}^0T}-\floor{z_k^0T}}}(\hat{m}_{i,j,l}^{\floor{z_k^0T}+1:\floor{zT}}-&\hat{m}_{i,j,l}^{\floor{z_k^0T}+1:\floor{z_{k+1}^0T}}) \\
 \,& \rightsquigarrow  \int_{[0,1]^2}\mathbb{D}_{k,i,j}(z,u,v)\,\dd h_l(u,v),
\end{split}
\end{equation}
so that
\(
\ssup\limits_{z \inn [z_{k}^0,z_{k+1}^0]} \hat{M}_k(z) \stackrel{d}{\rightarrow} \ssup\limits_{z \inn [z_{k}^0,z_{k+1}^0]}\mathbb{M}_k(z). 
\)
Since \(\{\mathbb{M}_k(z)\}_{k \eq 0}^{\ell^0}\) are independent and Gaussian, the claim follows by the continuous mapping theorem.

\renewcommand{\thelemma}{B.2}
\begin{lemma}\label{lem-auxB2}\textcolor[rgb]{1,1,1}{.}\vspace*{-.75cm}
\begin{enumerate}[label= \textnormal{(\arabic*)}]
\item\label{lema-auxB21} If there is no break in \([z_k^0,z_{k+1}^0]\) for all \(k \in \{0,1,\dots,\ell^0\}\), then
\[\normalfont
\mmax\limits_{0 \lleq k \lleq \ell}\left\vert\ssup\limits_{z \in [z_k^0,z_{k+1}^0]}\hat{M}(z_k^0,z_{k+1}^0;z)-\ssup\limits_{z \in [\hat{z}_k,\hat{z}_{k+1}]}\hat{M}(\hat{z}_k,\hat{z}_{k+1};z)\right\vert = O_p(1/\sqrt{T}).
\]
\item\label{lema-auxB22} If there is at least one break in \([z_k^0,z_{k+1}^0]\) for at least one \(k \in \{0,1,\dots,\ell\}\), then
\[\normalfont
\mmax\limits_{0 \lleq k \lleq \ell}\left\vert\ssup\limits_{z \in [z_k,z_{k+1}]}\hat{M}(z_k^0,z_{k+1}^0;z)-\ssup\limits_{z \in [\hat{z}_k,\hat{z}_{k+1}]}\hat{M}(\hat{z}_k,\hat{z}_{k+1};z)\right\vert = O_p(1).
\]
\end{enumerate}
\end{lemma}

\noindent \textbf{Proof of Lemma \ref{lem-auxB2}.} We assume, \textit{w.l.o.g.}, that \(\ell^0 = 1\); i.e., \(0 \eqqcolon z_0^0 < z_1^0 < z_2^0 \coloneqq 1\). Set \(\hat{M}_{1}(z) \coloneqq \hat{M}(0,z_1^0;z)\), \(\hat{M}_{\hat{z}_1}(z) \coloneqq \hat{M}(0,\hat{z}_{1};z)\) and define
\[
Z_{1}(\hat{z}_1) \coloneqq \ssup\limits_{z \inn [0,z_1^0]} \hat{M}_{1}(z) - \ssup\limits_{z \inn [0,\hat{z}_{1}]} \hat{M}_{\hat{z}_1}(z).
\] 
Next, define \(D_T \coloneqq \{z: z \in [0,1],\, |z - z_1| \leq c_0/T\}\) for some \(c_0 \in (0,\infty)\) and note that for any \(\delta > 0\) and \(\epsilon > 0\)
\begin{equation}
\begin{split}
\PP(|Z_{1}(\hat{z}_1)| > \delta) \leq \,& \PP(\hat{z}_1 \notin D_T) + \PP(\{|Z_{1}(\hat{z}_1)| > \delta\} \,\cap\, \{\hat{z}_1 \in D_T\})\\
\leq \,& \epsilon + \PP\bigg(\ssup\limits_{\zeta \inn D_T}|Z_{1}(\zeta)| > \delta\bigg).
\end{split}
\end{equation}
Hence, it suffices to show that \(|Z_{1}(\zeta)| = o_p(1)\) uniformly in \(\zeta \in D_T\). To see this, note that
\begin{equation}
\begin{split}
\left\vert\ssup\limits_{z \inn [0,z_1]} \hat{M}_{1}(z) - \ssup\limits_{z \inn [0,\zeta]} \hat{M}_{\zeta}(z) \right\vert
\leq \,& \ssup\limits_{z \inn [0,z_{1}]} \left\vert\hat{M}_{1}(z) - \hat{M}_{\zeta}(z) \right\vert  \\
\,& +\left\vert \ssup\limits_{z \inn [0,z_1]} \hat{M}_{\zeta}(z) - \ssup\limits_{z \inn [0,\zeta]} \hat{M}_{\zeta}(z)  \right\vert \coloneqq A(\zeta) + B(\zeta),
\end{split}
\end{equation}
say. Begin with \(A(\zeta)\). \textbf{Suppose, \textit{w.l.o.g.}, that \(\zeta > z_1\)}. Consider
\begin{equation}
\begin{split}
\left\vert\hat{M}_{1}(z) - \hat{M}_{\zeta}(z) \right\vert = \,& \frac{\floor{zT}}{\sqrt{\floor{\zeta T}}}\left\vert \sqrt{\frac{\floor{\zeta T}}{\floor{z_1 T}}}\norm{\hat{\mb}^{1:\floor{zT}}-\hat{\mb}^{1:\floor{z_1 T}}}_2-\norm{\hat{\mb}^{1:\floor{zT}}-\hat{\mb}^{1:\floor{\zeta T}}}_2\right\vert \\
\leq \,&  \sqrt{1-\frac{\floor{z_1 T}}{\floor{\zeta T}}} \hat{M}_{1}(z)  +  \frac{\floor{zT}}{\sqrt{\floor{\zeta T}}}A_1(\zeta),
\end{split}
\end{equation}
where \(A_1(\zeta)\coloneqq \norm{\hat{\mb}^{1:\floor{\zeta T}}-\hat{\mb}^{1:\floor{z_1 T}}}_2\); for the inequality, we expanded the term in absolute values on the right-hand side of the equality with \(\norm{\hat{\mb}^{1:\floor{zT}}-\hat{\mb}^{1:\floor{z_1 T}}}_2\), we used the (reverse) triangle inequality and \(\sqrt{u}-\sqrt{v} \leq \sqrt{u-v}\) for \(u \geq v \geq 0\). It follows that \(\ssup\limits_{z \inn [0,z_1]} \hat{M}_{1}(z) = O_p(1)\) if there is no break in \([0,z_1]\) or \(\ssup\limits_{z \inn [0,z_1]} \hat{M}_{1}(z)/\sqrt{T} = O_p(1)\) if there is at least one break in  \([0,z_1]\). Turning to \(A_1(\zeta)\), the triangle inequality and \(\norm{\vb}_2 \leq n\mmax_{1 \leq i \leq n} |v_i|\) for any \(\vb \in \mathbb{R}^n\) yields 
\begin{equation}\nn
\begin{split}
A_1(\zeta)
\leq \,& p\mmax_{1 \lleq l \lleq H}\mmax_{1 \lleq g \lleq G}\mmax\limits_{(i,j) \inn \mathcal{G}_g}\bigg[\frac{1}{\floor{\zeta T}}\sum_{t \eq \floor{z_1 T}}^{\floor{\zeta T}}\left\vert h_l(\hat{U}_{i,t}^{1:\floor{\zeta T}},\hat{U}_{j,t}^{1:\floor{\zeta T}})\right\vert \\
\,&\hspace*{.25cm} + \left(\frac{\floor{\zeta T}}{\floor{z_1 T}}-1\right) \frac{1}{\floor{z_1 T}}\sum_{t \eq 1}^{\floor{z_1 T}}\left\vert h_l(\hat{U}_{i,t}^{1:\floor{\zeta T}},\hat{U}_{j,t}^{1:\floor{\zeta T}})\right\vert \\
\,&\hspace*{.25cm} + \frac{1}{\floor{z_1 T}}\sum_{t \eq 1}^{\floor{z_1 T}}\left\vert h_l(\hat{U}_{i,t}^{1:\floor{\zeta T}},\hat{U}_{j,t}^{1:\floor{\zeta T}}) -h_l(\hat{U}_{i,t}^{1:\floor{z_1 T}},\hat{U}_{j,t}^{1:\floor{z_1 T}})\right\vert\bigg]
\end{split}
\end{equation}
say. It follows from assumption \ref{A-Lipsch} that \(\mmax_{1 \lleq l \lleq H}\ssup\limits_{(u,v) \inn [0,1]^2}|h_l(u,v)| \leq c_0\) for some \(c_0 \in (0,\infty)\). Hence the first to summands on the preceding display are \(O(1/T)\) uniformly in \(\zeta \in D_T\). Turning to the third summand, note that, by Assumption \ref{A-lipschitz}, one gets for some \(c_0 \in (0,\infty)\)
\begin{equation}
\begin{split}
\frac{1}{\floor{z_1 T}}\sum_{t \eq 1}^{\floor{z_1 T}}\Big\vert h_l(\hat{U}_{i,t}^{1:\floor{\zeta T}}&,\hat{U}_{j,t}^{1:\floor{\zeta T}}) -h_l(\hat{U}_{i,t}^{1:\floor{z_1 T}},\hat{U}_{j,t}^{1:\floor{z_1 T}}) \Big\vert \\
\,&\leq 2c_0\mmax\limits_{k \inn \{i,j\}}\ssup_{x \inn \mathbb{R}}\left\vert\hat{F}_k^{1:\floor{\zeta T}}(x)-\hat{F}_k^{1:\floor{z_1 T}}(x)\right\vert = O_p(1/T).
\end{split}
\end{equation}
Putting the above together yields, uniformly in \(\zeta \in D_T\), \(A(\zeta) = O_p(1/\sqrt{T})\) if there is no break and  \(A(\zeta) = O_p(1)\) if there is at least one break in \([0,1]\). Turning to \(B(\zeta)\), a simple calculation reveals that
\[
B(\zeta) \leq \ssup\limits_{z \inn [z_1,\zeta]}\left\vert \hat{M}_{\zeta}(z) -\hat{M}_{\zeta}(z_1)\right\vert \leq \ssup\limits_{z \inn [z_1,\zeta]} \frac{\floor{z T}}{\sqrt{\zeta T}}\norm{\hat{\mb}^{1:\floor{zT}}-\hat{\mb}^{1:\floor{z_1 T}}}_2 = O_p(1/\sqrt{T}),
\]
noting that there is no nontrivial break in \([z_1,\zeta]\).

\subsection{Additional Results for the Information Criterion}\label{ICproof}
Adopting the notation of Remark \ref{rem: infocrit} and using an idea from \cite{andrews1999consistent}, we introduce the set of selection vectors
\[
\mathcal{C}_{L} \coloneqq \left\{\cb = (c_1,\dots,c_{L})' \in \mathbb{R}^{L} \mid c_j \in \{0,1\},\,1 \leq j \leq L \right\}.
\]
Recall from Remark \ref{rem: infocrit} that we define for any \(\cb \in \mathcal{C}_L\) our break-point `information criterion' as
\begin{equation}\label{BPIC}
\textsf{IC}_T(\cb) \coloneqq \hat{S}^2(\cb) + h(|\cb|_1)\kappa_T, 
\end{equation}
where \(\hat{S}(\cb) \coloneqq  \hat{S}(\kb_c)\), with
\begin{equation}\nn
\begin{split}
\hat{S}(\kb_c)   \coloneqq \,&\mmax\limits_{1 \lleq j \lleq |\cb|_1+1}\mmax\limits_{k_{c,j-1} \,<\, t \lleq k_{c,j}}\hat{S}(k_{c,j-1},k_{c,j};t)\\
\hat{S}(k_{c,j-1},k_{c,j};t)\coloneqq \,&\frac{t-k_{c,j-1}}{\sqrt{k_{c,j}-k_{c,j-1}}}\norm{\hat{\mb}^{k_{c,j-1}+1:t}-\hat{\mb}^{k_{c,j-1}+1:k_{c,j}}}_2,\;\;\hat{k}_{0,c} \coloneqq 1,\,\hat{k}_{|\cb|_1+1,c} \coloneqq T,
\end{split}
\end{equation}
while \(h(\cdot)\) and the penalty \(\kappa_T\) are defined below.
\renewcommand{\theassumption}{IC}
\begin{assumption} 
\text{a)} The function \(h: \mathbb{R}_+ \rightarrow \mathbb{R}_+\) is strictly increasing. \text{b)} \(\kappa_T \rightarrow \infty\) and \(\kappa_T = o(T)\).
\end{assumption}
Assumption is the same as in \cite{andrews1999consistent} and \cite{luandrews2001}. Finally, we define the estimator \(\hat{\cb}\) \textit{via}
\[
\hat{\cb} \coloneqq \argmin_{\cb \inn \mathcal{C}_L}\textsf{IC}_T(\cb).
\]
Our objective here is twofold. \textit{Firstly}, we want to obtain at least as many break points as to ensure that the corresponding intervals \(([\hat{k}_{j-1,c},\hat{k}_{j,c}])_{j \eq 1}^{|\cb|_1+1}\), do not contain further nontrivial break points. If this is indeed the case, then Proposition \ref{prop:max_lim} reveals \(\hat{S}(\cb)/\sqrt{T} = o_p(1)\); on the other hand, if there is at least one nontrivial break in between \(([\hat{k}_{j-1,c},\hat{k}_{j,c}])_{j \eq 1}^{|\cb|_1+1}\), then, by Lemma \ref{lem-auxB2}, \(\hat{S}(\cb)/\sqrt{T} \stackrel{p}{\rightarrow} \textsf{const} \in (0,\infty)\). Thus our candidate \(\cb \in \mathcal{C}_L\) should, with probability approaching one (wp \(\rightarrow 1\)), satisfy \(\cb \in \mathcal{B}_L\),  where
\[
\mathcal{B}_{L} \coloneqq \left\{\cb \in \mathcal{C}_{L}  \mid \hat{S}(\cb)/\sqrt{T} = o_p(1)\right\}.
\]
\textit{Secondly}, provided some \(\cb \in \mathcal{B}_L\) is chosen, we would like to obtain the most parsimonious set of break points; i.e. \(\cb\) should fall wp \(\rightarrow 1\) into the set
\[
\mathcal{MB}_{L} \coloneqq \left\{\cb \in \mathcal{B}_{L}  \mid |\cb| \leq |\cb'|,\; \forall\, \cb' \in \mathcal{B}_{L}\right\}.
\]
We can show that \(\hat{\cb} \in \mathcal{MB}_{L}\) wp \(\rightarrow 1\), provided \(\ell_0 \leq L\). Since \(\mathcal{MB}_{L} = \{\cb_0\}\), it follows immediately that \(|\hat{\cb}|_1  \stackrel{p}{\rightarrow} \ell^0\). 
 
\begin{proposition}
 If \(\ell^0 \leq L\), then \((\hat{\kb}_{\hat{c}}/T,|\hat{\cb}|_1) \stackrel{p}{\longrightarrow} (\zb^0,\ell^0)\), where \(\zb^0 \coloneqq (z_{1}^0,\dots,z_{\ell^0}^0)'\); compare assumption \ref{A-break}.
\end{proposition}
 
\noindent \textbf{Proof.} The following mimics the proof in \cite{andrews1999consistent}. It follows readily that \(\hat{S}^2(\kb)/T = o_p(1)\) if there is no break and \(\hat{S}^2(\kb)/T \stackrel{p}{\rightarrow} \textsf{const} \in (0,\infty)\), otherwise. By definition of \eqref{BPIC}, one concludes \(\hat{\cb} \in \mathcal{B}_L\) wp \(\rightarrow 1\). If we take \(\cb_1, \cb_2 \in \mathcal{B}_L\) with \(\cb_1 \in \mathcal{MB}_{L}\) and \(\cb_2 \notin \mathcal{MB}_{L}\), then \(|\cb_1|_1 \leq |\cb_2|_1\) and \((h(|\cb_1|_1)-h(|\cb_2|_1))\kappa_T \rightarrow - \infty\). Therefore, \(\textsf{IC}_T(\cb_1) \leq \textsf{IC}_T(\cb_2)\) wp \(\rightarrow 1\) so that \(\hat{\cb} \in \mathcal{MB}_{L} = \{\cb_0\}\) wp \(\rightarrow 1\).

\section[Appendix B: Tables]{Tables} \label{app: Tables}
\setcounter{table}{0}
\numberwithin{table}{section}
\begin{table}[H]
\centering
\setlength{\tabcolsep}{2pt}
\footnotesize
\begin{tabular}{c|c|c|c|c|c|c|c|c|c|c}
\cline{1-11}
loadings & 0 breaks	& \multicolumn{2}{c|}{1 break}		& \multicolumn{3}{c|}{2 breaks} 		& \multicolumn{4}{c}{3 breaks} \\
&$[1,T]$&\multicolumn{1}{c}{$[1,t_1]$}&$(t_1,T]$&\multicolumn{1}{c}{$[1,0.4T]$}&\multicolumn{1}{c}{$(0.4T,0.6T]$}&$(0.6T,T]$&\multicolumn{1}{c}{$[1 ,0.3T]$}&\multicolumn{1}{c}{$(0.3T,0.5T]$}&\multicolumn{1}{c}{$(0.5T,0.7T]$}&$(0.7T,T]$\\		
\cline{1-11}
$\beta_{1,t}$ & 1 & 1 & 0.5 & 1 & 0.5 & 1 & 0.7 & 1 & 0.4 & 0.7\\
$\beta_{2,t}$ & 0.5 & 0.5 & 1 & 0.5 & 1 & 0.5 & 0.7 & 0.4 & 1 & 0.7  \\
$\beta_{3,t}$ & 1 & 1 & 1 & 1 & 1 & 1 & 0.7 & 0.7 & 0.7 & 0.7\\
$\beta_{4,t}$ & 1.5 & 1.5 & 0.5 & 1.5 & 0.5 & 1.5 & 0.7 & 1.2 & 0.2 & 0.7\\
\hhline{=|=|=|=|=|=|=|=|=|=|=}
$\rho_1$ & 0.4469 & 0.4469 & 0.2006 & 0.4469 & 0.2006 & 0.4469 & 0.3074 & 0.4469 & 0.1459 & 0.3074\\
$\rho_2$ & 0.2006 & 0.2006 & 0.4469 & 0.2006 & 0.4469 & 0.2006 & 0.3074 & 0.1459 & 0.4469 & 0.3074\\
$\rho_3$ & 0.4469 & 0.4469 & 0.4469 & 0.4469 & 0.4469 & 0.4469 & 0.3074 & 0.3074 & 0.3074 & 0.3074\\
$\rho_4$ & 0.6132 & 0.6132 & 0.2006 & 0.6132 & 0.2006 & 0.6132 & 0.3074 & 0.5212 & 0.0465 & 0.3074\\
\cline{1-11}
$\bar{\rho}$ & 0.4269 & 0.4269 & 0.3238 & 0.4269 & 0.3238 & 0.4269 & 0.3074 & 0.3554 & 0.2367 & 0.3074\\
\end{tabular}
\captionof{table}{loadings from the DGP \eqref{MCfactor} with $4$ groups in the corresponding interval. $\rho_i$ contains the cross-sectional average of Spearman's $\rho$ in group $i$ averaged over 10000 simulations. $\bar{\rho}$ is the average over $\rho_i$}
\label{tab: loadings}
\end{table}

\newpage
\begin{table}[h]
\centering
\begin{tabular}{|c|c|ccc|c|c|}
\cline{1-7}
\multirow{3}{*}{$BS$}		&0breaks & \multicolumn{3}{c|}{1break}& 2breaks& 3breaks \\
	     	&& $z_1$=0.15& $z_1$=0.5& $z_1$=0.85	 & $z_1$=0.4 & $z_1$=0.3 \\
	     	&&&&&$z_2$=0.6&$z_2$=0.5 $z_3$=0.7 \\
\cline{1-7}
-3		&	  &		&		& & & 0.043	\\
-2		&	  &		&		&& 0.354	& 0.013\\
-1		&	  & 0.000 & 0.000 & 0.037	  & 0.000	&0.322\\
0		&0.906& 0.600 & 0.931 & 0.711	  & 0.558&0.597\\
1		&0.085& 0.370 & 0.068 & 0.236   & 0.084&0.024\\
2		&0.008& 0.028 & 0.001 & 0.016	  & 0.004&0.001\\
3		&0.001& 0.002 & 0.000 & 0.000	  & 0.000&0.000\\
\cline{1-7}
\multicolumn{1}{|c}{\multirow{2}{*}{$WBS$}}		&\multicolumn{1}{c}{}& \multicolumn{3}{c}{}			 &\multicolumn{1}{c}{}&\\\multicolumn{1}{|c}{}&\multicolumn{1}{c}{}&&&\multicolumn{1}{c}{}&\multicolumn{1}{c}{}&\multicolumn{1}{c|}{}\\			
\cline{1-7}
-3		&	  &		&		&&  &0.008\\
-2		&	  &		&		&&	0.013 & 0.461\\
-1		&	  & 0.284 & 0.000 &	0.549 & 0.233 &0.430	\\
0		&0.902& 0.621 & 0.933 & 0.404 & 0.678&0.093\\
1 		&0.092& 0.086 & 0.065 & 0.043 & 0.075&0.008\\
2		&0.006& 0.009 & 0.002 & 0.004 & 0.001&0.000\\
3		&0.000& 0.000 & 0.000 & 0.000 & 0.000 &0.000\\
\cline{1-7}
\multicolumn{1}{|c}{\multirow{2}{*}{$WBS_{BS}$}}		&\multicolumn{1}{c}{}& \multicolumn{3}{c}{}			&\multicolumn{1}{c}{}& \\ \multicolumn{1}{|c}{}&\multicolumn{1}{c}{}&&&\multicolumn{1}{c}{}&\multicolumn{1}{c}{}&\multicolumn{1}{c|}{}\\		
\cline{1-7}
-3		&	  &		&		&	&	    &0.012\\
-2		&	  &		&		&& 0.016	&0.312\\
-1		&	  & 0.092 & 0.000 & 0.488 & 0.082 &0.517\\
0		&0.907& 0.753 & 0.925 & 0.443 & 0.882 &0.148\\
1 		&0.086& 0.143 & 0.073 & 0.065 & 0.076 &0.011\\
2		&0.007& 0.012 & 0.002 & 0.004 & 0.004 &0.000\\
3		&0.000& 0.000 & 0.000 & 0.000 & 0.000 &0.000\\
\cline{1-7}
\end{tabular}
\captionof{table}{Over- and underestimation frequencies of the amount of true change points $\hat{\ell}-\ell^0$ with $\alpha = 0.01$ and $T=500$}
\label{tab: over-/underestimation dependencies T=500 alpha=0.1}
\end{table}

\newpage
\begin{table}[h]
\centering
\begin{tabular}{|c|c|ccc|c|c|}
\cline{1-7}
\multirow{3}{*}{$BS$}		&0breaks & \multicolumn{3}{c|}{1break}& 2breaks& 3breaks \\
	     	&& $z_1$=0.15& $z_1$=0.5& $z_1$=0.85	 & $z_1$=0.4 & $z_1$=0.3 \\
	     	&&&&&$z_2$=0.6&$z_2$=0.5 $z_3$=0.7 \\
\cline{1-7}
-3		&	  &		&		& &  & 0.000	\\
-2		&	  &		&		&& 0.004	& 0.000\\
-1		&	  & 0.000 & 0.000 & 0.000	  & 0	&0.000\\
0		&0.958& 0.562 & 0.974 & 0.861	  & 0.936&0.968\\
1		&0.039& 0.427 & 0.026 & 0.137   & 0.060&0.032\\
2		&0.003& 0.011 & 0.000 & 0.002	  & 0.000&0.000\\
3		&0.000& 0.000 & 0.000 & 0.000	  & 0.000&0.000\\
\cline{1-7}
\multicolumn{1}{|c}{\multirow{2}{*}{$WBS$}}		&\multicolumn{1}{c}{}& \multicolumn{3}{c}{}			 &\multicolumn{1}{c}{}&\\\multicolumn{1}{|c}{}&\multicolumn{1}{c}{}&&&\multicolumn{1}{c}{}&\multicolumn{1}{c}{}&\multicolumn{1}{c|}{}\\			
\cline{1-7}
-3		&	  &		&		&&  &0.000\\
-2		&	  &		&		&&	0.000 & 0.005\\
-1		&	  & 0.017 & 0.000	& 0.058		 & 0.002 &0.146	\\
0		&0.958& 0.864 & 0.978 & 0.904	  & 0.972&0.827\\
1 		&0.042& 0.113 & 0.021 & 0.036   & 0.025&0.022\\
2		&0.000& 0.006 & 0.001 & 0.002	  & 0.001&0.000\\
3		&0.000& 0.000 & 0.000 & 0.000	  & 0.000 &0.000\\
\cline{1-7}
\multicolumn{1}{|c}{\multirow{2}{*}{$WBS_{BS}$}}		&\multicolumn{1}{c}{}& \multicolumn{3}{c}{}			&\multicolumn{1}{c}{}& \\ \multicolumn{1}{|c}{}&\multicolumn{1}{c}{}&&&\multicolumn{1}{c}{}&\multicolumn{1}{c}{}&\multicolumn{1}{c|}{}\\		
\cline{1-7}
-3		&	  &		&		&	&	    &0.000\\
-2		&	  &		&		&& 0.000 &0.000\\
-1		&	  & 0.000 & 0.000 & 0.000	&  0.000&0.026\\
0		&0.974& 0.743 & 0.973 & 0.943	&  0.979&0.944\\
1 		&0.023& 0.250 & 0.027 & 0.055 &  0.020&0.030\\
2		&0.003& 0.007 & 0.000 & 0.002	& 0.001&0.000\\
3		&0.000& 0.000 & 0.000 & 0.000	& 0.000&0.000\\
\cline{1-7}
\end{tabular}
\captionof{table}{Over- and underestimation frequencies of the amount of true change points $\hat{\ell}-\ell^0$ with $\alpha = 0.0333$ and $T=1500$}
\label{tab: over-/underestimation dependencies T=1500 alpha=0.05}
\end{table}

\newpage
\begin{table}[h]
\centering
\begin{tabular}{|c|c|ccc|c|c|}
\cline{1-7}
\multirow{3}{*}{$BS$}		&0breaks & \multicolumn{3}{c|}{1break}& 2breaks& 3breaks \\
	     	&& $z_1$=0.15& $z_1$=0.5& $z_1$=0.85	 & $z_1$=0.4 & $z_1$=0.3 \\
	     	&&&&&$z_2$=0.6&$z_2$=0.5 $z_3$=0.7 \\
\cline{1-7}
-3		&	  &		&		& & & 0.000	\\
-2		&	  &		&		&& 0.035	& 0.000\\
-1		&	  & 0.000 & 0.000 & 0.000 & 0.000	&0.064\\
0		&0.944& 0.627 & 0.959 & 0.800 & 0.888&0.896\\
1		&0.055& 0.357 & 0.039 & 0.196 & 0.077&0.039\\
2		&0.001& 0.015 & 0.002 & 0.004 & 0.000&0.001\\
3		&0.000& 0.001 & 0.000 & 0.000 & 0.000&0.000\\
\cline{1-7}
\multicolumn{1}{|c}{\multirow{2}{*}{$WBS$}}		&\multicolumn{1}{c}{}& \multicolumn{3}{c}{}			 &\multicolumn{1}{c}{}&\\\multicolumn{1}{|c}{}&\multicolumn{1}{c}{}&&&\multicolumn{1}{c}{}&\multicolumn{1}{c}{}&\multicolumn{1}{c|}{}\\			
\cline{1-7}
-3		&	  &		&		&&  &0.000\\
-2		&	  &		&		&&	0.000 & 0.108\\
-1		&	  & 0.060 & 0.000 & 0.139 & 0.007 &0.509	\\
0		&0.950& 0.837 & 0.956 & 0.809 & 0.945&0.371\\
1 		&0.047& 0.093 & 0.042 & 0.050 & 0.046&0.012\\
2		&0.003& 0.010 & 0.002 & 0.002 & 0.002&0.000\\
3		&0.000& 0.000 & 0.000 & 0.000 & 0.000 &0.000\\
\cline{1-7}
\multicolumn{1}{|c}{\multirow{2}{*}{$WBS_{BS}$}}		&\multicolumn{1}{c}{}& \multicolumn{3}{c}{}			&\multicolumn{1}{c}{}& \\ \multicolumn{1}{|c}{}&\multicolumn{1}{c}{}&&&\multicolumn{1}{c}{}&\multicolumn{1}{c}{}&\multicolumn{1}{c|}{}\\		
\cline{1-7}
-3		&	  &		&		&	&	    &0.000\\
-2		&	  &		&		&& 0.000	&0.021\\
-1		&	  & 0.000 & 0.000 & 0.019 & 0.001 &0.379\\
0		&0.950& 0.825 & 0.951 & 0.898 & 0.943 &0.568\\
1 		&0.048& 0.168 & 0.044 & 0.078 & 0.053 &0.032\\
2		&0.002& 0.006 & 0.005 & 0.005 & 0.003 &0.000\\
3		&0.000& 0.001 & 0.000 & 0.000 & 0.000 &0.000\\
\cline{1-7}
\end{tabular}
\captionof{table}{Over- and underestimation frequencies of the amount of true change points $\hat{\ell}-\ell^0$ for filtered data with $\alpha = 0.05$ and $T=1000$}
\label{tab: over-/underestimation dependencies T=1000 alpha=0.05, filtered}
\end{table}

\end{appendices}


\newpage
\addcontentsline{toc}{chapter}{References}
\bibliographystyle{apalike} 
\bibliography{References} 


\end{document}